\documentclass[reqno, eucal]{amsart}

\usepackage[mathscr]{eucal} 
\usepackage[dvips]{graphicx}
\usepackage[dvips]{color}
\usepackage{amsmath,amsfonts,amssymb,amsthm,amscd}
\usepackage{epic}
\usepackage{eepic}
\usepackage{longtable}
\usepackage{array}

\newtheorem{theorem}{Theorem}[section]

\theoremstyle{definition}

\newtheorem{example}[theorem]{Example}
\newtheorem{remark}[theorem]{Remark}

\newcommand{\Z}{{\mathbb Z}}

\newcommand{\C}{{\mathbb C}}

\newcommand{\sk}{{\rm{\bf s}}} 
\newcommand{\ok}{{\rm{\bf k}}}
\newcommand{\OK}{{\rm{\bf K}}}
\newcommand{\am}{{\rm{\bf a}}^{\!-} }
\newcommand{\ap}{{\rm{\bf a}}^{\!+} }
\newcommand{\apm}{{\rm{\bf a}}^{\!\pm} }
\newcommand{\amp}{{\rm{\bf a}}^{\!\mp} }

\newcommand{\Am}{{\rm{\bf A}}^{\!-} }
\newcommand{\Ap}{{\rm{\bf A}}^{\!+} }
\newcommand{\Apm}{{\rm{\bf A}}^{\!\pm} }
\newcommand{\Amp}{{\rm{\bf A}}^{\!\mp} }
\newcommand{\ichi}{1}
\newcommand{\hf}{{\scriptstyle \frac{1}{2}}}
\newcommand{\thf}{{\scriptstyle \frac{3}{2}}}

\newcommand{\Rm}{\mathscr{R}}

\newcommand{\Fm}{\mathscr{F}}
\newcommand{\Vb}{{\bf V}}
\newcommand{\alb}{\boldsymbol{\alpha}}
\newcommand{\beb}{\boldsymbol{\beta}}
\newcommand{\gab}{\boldsymbol{\gamma}}
\newcommand{\deb}{\boldsymbol{\delta}}

\newcommand{\lab}{\boldsymbol{\lambda}}
\newcommand{\mub}{\boldsymbol{\mu}}
\newcommand{\nub}{\boldsymbol{\nu}}

\begin{document}

\title[$G_2$ reflection equation]
{Matrix product solutions to the $\boldsymbol{G_2}$ reflection equation}

\author{Atsuo Kuniba}
\email{atsuo.s.kuniba@gmail.com}
\address{Institute of Physics, 
University of Tokyo, Komaba, Tokyo 153-8902, Japan}

\maketitle

\vspace{0.5cm}
\begin{center}{\bf Abstract}
\end{center}

We study the $G_2$ reflection equation
for the three particles in $1+1$ dimension 
that undergo a special scattering/reflections 
described by the Pappus theorem. 
It is a sixth order equation and 
serves as a natural $G_2$ analogue of the Yang-Baxter and the reflection equations 
corresponding to the cubic and the quartic Coxeter relations 
of type $A$ and $BC$, respectively.
We construct matrix product solutions to the $G_2$ reflection equation 
by exploiting a connection to the representation theory 
of the quantized coordinate ring $A_q(G_2)$.
\vspace{0.4cm}

\section{Introduction}

The Yang-Baxter and the reflection equations
\begin{equation}\label{asocia}
\begin{split}
R_{12}(\alpha_1)R_{13}(\alpha_1+\alpha_2)R_{23}(\alpha_2)
&= R_{23}(\alpha_2)R_{13}(\alpha_1+\alpha_2)R_{12}(\alpha_1),\\
R_{12}(\alpha_1)K_2(\alpha_1+\alpha_2)
R_{21}(\alpha_1+2\alpha_2)K_1(\alpha_2)
&=K_1(\alpha_2)R_{12}(\alpha_1+2\alpha_2)
K_2(\alpha_1+\alpha_2)R_{21}(\alpha_1)
\end{split}
\end{equation}
are the fundamental structure in quantum integrable systems 
in the bulk \cite{Bax} and at the boundary \cite{Ch,Sk,Kul}.
They are the Yang-Baxterizations (spectral parameter dependent versions)
of the cubic and the quartic Coxeter relations for the 
simple reflections $s_1, s_2$ of the root systems of $A_2$ and $B_2/C_2$:
\begin{align*}
s_1s_2s_1 &= s_2s_1s_2,\quad \quad
\Delta_+ =\{\alpha_1,\alpha_1+\alpha_2, \alpha_2\},
\\
s_1s_2s_1s_2&= s_2s_1s_2 s_1,\quad\, 
\Delta_+ =\{\alpha_1,\alpha_1+\alpha_2, \alpha_1+2\alpha_2, \alpha_2\}.
\end{align*}
Here $\alpha_1, \alpha_2$ are the simple roots and 
$\Delta_+$ denotes the set of positive roots which formally 
correspond to the spectral parameters.
They are so ordered that the $k$th one from the left is 
$s_{i_1}\cdots s_{i_{k-1}}(\alpha_{i_k})$
with $i_k=1$ ($k$: odd) and $i_k=2$ ($k$: even).
For simplicity we assume $R \in \mathrm{End}(\Vb \otimes \Vb)$ and 
$K \in \mathrm{End}\, \Vb$ for some vector space $\Vb$.

It is natural and by now classic 
to extend the `factorization' condition like (\ref{asocia}) 
to more general root systems \cite{Ch}.
In this paper we study the $G_2$ case of the form
\begin{equation}\label{reika0}
\begin{split}
&R_{12}(\alpha_1)G_{132}(\alpha_1+\alpha_2)R_{23}(2\alpha_1+3\alpha_2)
G_{213}(\alpha_1+2\alpha_2)R_{31}(\alpha_1+3\alpha_2)G_{321}(\alpha_2)\\
&=G_{231}(\alpha_2)R_{13}(\alpha_1+3\alpha_2)G_{123}(\alpha_1+2\alpha_2)
R_{32}(2\alpha_1+3\alpha_2)G_{312}(\alpha_1+\alpha_2)R_{21}(\alpha_1),
\end{split}
\end{equation}
where $R$ is a solution to the Yang-Baxter equation by itself and 
$G \in \mathrm{End}(\Vb \otimes \Vb \otimes \Vb)$ is the 
characteristic operator in the $G_2$ theory.
It is a Yang-Baxterization of the $G_2$ Coxeter relation
$s_1s_2s_1s_2s_1s_2= s_2s_1s_2 s_1s_2s_1$ with the 
spectral parameters corresponding to the positive roots. 
See (\ref{theta})--(\ref{mirk}).

Although the equation (\ref{reika0}) 
was not written down explicitly in \cite{Ch},
it was explained to the author by 
Cherednik \cite{Ch3} that the $G_2$ factorization condition 
is depicted by a three particle scattering diagram corresponding to (\ref{reika0}) 
and it is related to the geometry of the Desargues-Pappus theorem\footnote{
This is partly described in \cite[p982]{Ch} 
and will be detailed in Section \ref{ss:pap}.}. 
The equation (\ref{reika0}) for generic symbols $R$ and $G$ without assuming a 
tensor structure on their representation space (i.e.~without indices) has appeared as a 
defining relation of the {\em root algebra} of type $G_2$ \cite[Sec.2]{Ch2}.
In this paper we call (\ref{reika0}) 
{\em the $G_2$ reflection equation} for simplicity.

The purpose of this paper is to construct families of solutions 
to the $G_2$ reflection equation with
$\Vb = (\C^2)^{\otimes n}$ for any positive integer $n$.
Our approach is based on the 3 dimensional (3D) integrability 
developed in \cite{BS, KS, KOS1,KOS2} 
for the Yang-Baxter equation and in \cite{KP} for the reflection equation.
The most essential idea of it is to embark on a {\em quantization} or 
a {\em 3D version} of the $G_2$ reflection equation.
We introduce the {\em quantized $G_2$ reflection equation} 
\begin{align}\label{aimi}
(L_{12}J_{132}L_{23}J_{213}L_{31}J_{321}) \circ \Fm
=\Fm \circ (J_{231}L_{13}J_{123}L_{32}J_{312}L_{21}),
\end{align}
which is a $G_2$ reflection equation (without spectral parameters) 
up to conjugation by a certain 
operator $\Fm$ acting on an auxiliary $q$-boson Fock space.
Our finding (Theorem \ref{th:sol}) is that with a suitable choice of the 
quantized scattering amplitude $L$ and $J$,
(\ref{aimi}) coincides exactly with the intertwining relation 
\cite[eq.(28)]{KOY} of the $A_q(G_2)$ modules 
labeled by the longest element of the Weyl group \cite{So2}.
The $\Fm$ corresponds to the intertwiner.
Here $A_q(\mathfrak{g})$, 
for a finite dimensional classical simple Lie algebra $\mathfrak{g}$ in general,
denotes a Hopf subalgebra of the dual $U_q(\mathfrak{g})^\ast$
called {\em quantized coordinate ring}.  
It has been studied from a variety of aspects.
See \cite{D,RTF,So2,NYM, Kas93, Sasaki, GLS, KOY, S,T} for example.

In short, we obtain a solution to the quantized $G_2$ reflection equation (\ref{aimi}).
It offers a bonus;  the equation/solution can be concatenated  
along the $q$-boson Fock space for arbitrary $n$ times.
The piled $n$ layers of the $1+1$ dimensional scattering diagrams 
can be viewed as a 3D lattice system in which adjacent layers 
may be interchanged locally according to (\ref{aimi}) without changing the 
total statistical weight, a feature roughly referred to as 3D integrability.
Anyway, 
to the $n$-concatenation of the quantized $G_2$ reflection equation, 
one can insert the spectral parameters and 
evaluate the intertwiner $\Fm$ away appropriately.
It brings us back to the original $G_2$ reflection equation, thereby 
producing a solution to it for each $n$.
Actually there are two such recipes called trace reduction and 
boundary vector reduction.
They lead to the solutions 
$(R^{\mathrm{tr}}(z), G^{\mathrm{tr}}(z))$ and 
$(R^{\mathrm{bv}(z)}, G^{\mathrm{bv}}(z))$\footnote{
This latter solution assumes the
relation (\ref{syki}) yet to be proved.}, respectively.
By the construction they possess the matrix product structure
containing $n$-product of $L$'s or $J$'s\footnote{
$\varrho^{\mathrm{tr}}(z), \varrho^{\mathrm{bv}}(z),
\kappa^{\mathrm{tr}}(z), \kappa^{\mathrm{bv}}(z)$ are scalars
given in (\ref{askS}) and (\ref{prpr}).}
\begin{equation}\label{miho}
\begin{split}
R^{\mathrm{tr}}(z) 
&= \varrho^{\mathrm{tr}}(z)\mathrm{Tr}(z^{\bf h}L \cdots L),\qquad\;\;
G^{\mathrm{tr}}(z) 
= \kappa^{\mathrm{tr}}(z)\mathrm{Tr}(z^{\bf h}J \cdots J),\\
R^{\mathrm{bv}}(z) 
&= \varrho^{\mathrm{bv}}(z)
\langle \xi | z^{\bf h}L \cdots L | \xi \rangle ,\qquad
G^{\mathrm{bv}}(z) 
= \kappa^{\mathrm{bv}}(z)
\langle \xi | z^{\bf h}J \cdots J | \xi \rangle,
\end{split}
\end{equation}
where the trace and the sandwich $\langle \xi | (\cdots)  | \xi \rangle$
are taken over a $q$-boson Fock space. 
The detail will be explained in later sections.
The solutions are trigonometric in the spectral parameter\footnote{ ``Trigonometric"
means rational in $z$ in (\ref{miho}) which corresponds to the exponential  
of the spectral parameters in (\ref{reika0}).}. 
In fact $R^{\mathrm{tr}}(z)$ and $R^{\mathrm{bv}}(z)$ 
turn out to be the quantum $R$ matrices \cite{D,Ji}
for the antisymmetric tensor representations of $U_p(A^{(1)}_{n-1})$
and the spin representation of $U_p(D^{(2)}_{n+1})$ with $p^2=-q^{-3}$.
This part of the results is contained in the earlier works \cite{BS, KS}.

This paper may be viewed as a continuation of \cite{BS,KS} and \cite{KP}
where analogous results were obtained  
for the Yang-Baxter and the reflection equations, respectively.
To explore applications of the $G_2$ reflection equation is a future problem.
For instance to architect commuting transfer matrices 
based on the $G_2$ reflection is an interesting issue.

The paper is organized as follows.
In Section \ref{sec:g2} we explain the 
interpretation of the $G_2$ reflection equation in terms of 
a special three particle scattering following \cite{Ch, Ch3}.
The characteristic feature is the operator $G$ 
which encodes the simultaneous reflection of 
one of the particles at the boundary and scattering of the other two. 
The world-lines of these particles form a configuration 
matching the classical Pappus theorem.

In Section \ref{sec:qgre}
we formulate the quantized $G_2$ reflection equation by promoting 
$R$ and $G$ in (\ref{reika0}) to the $q$-boson valued $L$ and $J$.
The $L$ matrix (\ref{Lop}) appeared first in \cite{BS}.
The $q$-boson valued amplitude $J$ (\ref{Jdef})--(\ref{spd}) 
has been designed deliberately to validate Theorem \ref{th:sol}.
It does {\em not} split into the product of operators 
representing the single particle reflection and the two particle scattering.
See (\ref{gkkun}). 

In Section \ref{sec:rk}
after recalling basic facts on the representation theory of 
$A_q(G_2)$ \cite{So2},  we state our key observation in Theorem \ref{th:sol}. 
It identifies the quantized $G_2$ reflection equation with
the intertwining relation between certain $A_q(G_2)$ modules.

In Section \ref{sec:ybe}
we review the reduction of the tetrahedron equation (cf.~\cite{Zam80})
to the Yang-Baxter equation following \cite{BS,KS,KP}.
This construction has been illustrated 
in many literatures recently, e.g.~\cite{KOS1,KOS2},
so we keep the description brief.
A slightly more detailed exposition is available in \cite[App.B]{KP}.

In Section \ref{sec:re} we explain that the analogous reduction 
works perfectly also for the 
quantized $G_2$ reflection equation.
They lead to two families of solutions 
$(R^{\mathrm{tr}}(z), G^{\mathrm{tr}}(z))$ 
and $(R^{\mathrm{bv}}(z), G^{\mathrm{bv}}(z))$
to the $G_2$ reflection equation (\ref{reika0}), 
where the latter is yet based on the conjectural relation (\ref{syki}).
The role of the intertwiner $\Fm$ is curious.
Although it is complicated and no closed formula is known, 
it does not give rise to a difficulty since the reduction procedure just eliminates it.
Nevertheless $\Fm$ essentially controls the construction behind the scene 
in that it specifies precisely how the $L$ and $J$ are to be combined, 
how the spectral parameters should be arranged 
and what kind of boundary vectors are acceptable.
These are essential legacy of $\Fm$.

Section \ref{sec:end} is a summary.
Appendix \ref{app:ind} describes the precise correspondence between 
the quantized $G_2$ reflection equation and the 
intertwining relation (\ref{kkna}) of the $A_q(G_2)$ modules.
Appendix \ref{app:ex} contains explicit forms of 
$(R^{\mathrm{tr}}(z), G^{\mathrm{tr}}(z))$ 
and $(R^{\mathrm{bv}}(z), G^{\mathrm{bv}}(z))$
for small $n$.

Throughout the paper we assume that $q$ is generic and 
use the following notation:
\begin{align*}
&(z;q)_m = \prod_{k=1}^m(1-z q^{k-1}),\qquad
(q)_m = (q; q)_m,\\
&\theta(\text{true})=1,\;\;\theta(\text{false}) = 0,\quad
{\bf e}_j = (0,\ldots,0,\overset{j}{1},0,\ldots, 0) \in \Z^n\;(1 \le j \le n).
\end{align*}

\section{$G_2$ reflection equation for three particle scattering}\label{sec:g2}

\subsection{\mathversion{bold}The $G_2$ reflection equation}

Let $\Vb$ be a vector space and consider the operators 
\begin{align}\label{RG}
R(z) \in \mathrm{End}(\Vb \otimes \Vb),\qquad
G(z) \in \mathrm{End}(\Vb \otimes \Vb \otimes \Vb)
\end{align}
depending on the spectral parameter $z$.
We assume that $R(z)$ satisfies the Yang-Baxter equation:
\begin{align}\label{ybe}
R_{12}(x)R_{13}(xy)R_{23}(y) = R_{23}(y)R_{13}(xy)R_{12}(x)
\in \mathrm{End}(\Vb \otimes \Vb \otimes \Vb).
\end{align}
By the $G_2$ reflection equation we mean the following  
in $\mathrm{End}(\Vb \otimes \Vb \otimes \Vb)$:
\begin{equation}\label{reika}
\begin{split}
&R_{12}(x)G_{132}(xy)R_{23}(x^2y^3)
G_{213}(xy^2)R_{31}(xy^3)G_{321}(y)\\
&=G_{231}(y)R_{13}(xy^3)G_{123}(xy^2)
R_{32}(x^2y^3)G_{312}(xy)R_{21}(x).
\end{split}
\end{equation}
To explain the notation, write temporarily as
$R(z) = \sum r^{(1)}_l \otimes r^{(2)}_l$ and 
$G(z) = \sum g^{(1)}_l \otimes g^{(2)}_l \otimes g^{(3)}_l$ 
with some sums over $l$\footnote{ 
Although these expansions do not specify $r^{(a)}_l, g^{(a)}_l$ uniquely,
it suffices to make (\ref{ruby}) unambiguous.}. 
Then 
\begin{equation}\label{ruby}
\begin{split}
R_{12}(z) &= \sum r^{(1)}_l \otimes r^{(2)}_l \otimes \ichi,\quad
R_{13}(z) = \sum r^{(1)}_l \otimes \ichi \otimes r^{(2)}_l,\quad
R_{23}(z) = \sum \ichi \otimes  r^{(1)}_l \otimes r^{(2)}_l,\\
R_{21}(z) &= \sum r^{(2)}_l \otimes r^{(1)}_l \otimes \ichi,\quad
R_{31}(z) = \sum r^{(2)}_l \otimes \ichi \otimes r^{(1)}_l,\quad
R_{32}(z) = \sum \ichi \otimes  r^{(2)}_l \otimes r^{(1)}_l,\\
G_{ijk}(z) &= \sum g^{(i)}_l \otimes g^{(j)}_l \otimes g^{(k)}_l.
\end{split}
\end{equation}

\subsection{Scattering diagram; Pappus configuration}\label{ss:pap}

Let us describe the special three particle scattering
related to the $G_2$ reflection equation.
This is due to \cite{Ch,Ch3}.
Consider the three particles 1,2,3 
coming from A$_1$,A$_2$,A$_3$ and being  
reflected by the boundary at O$_1$, O$_2$, O$_3$, respectively. 
See Figure \ref{RS1}.
The bottom horizontal line is the boundary which may also  
be viewed as the time axis.
The vertical direction corresponds to the 1D space. 
Each line carries $\Vb$
which specifies an internal degrees of the freedom of a particle.
So a three particle state at a time is described by an element in 
$\Vb \otimes \Vb \otimes \Vb$.

\setlength\unitlength{0.5mm}
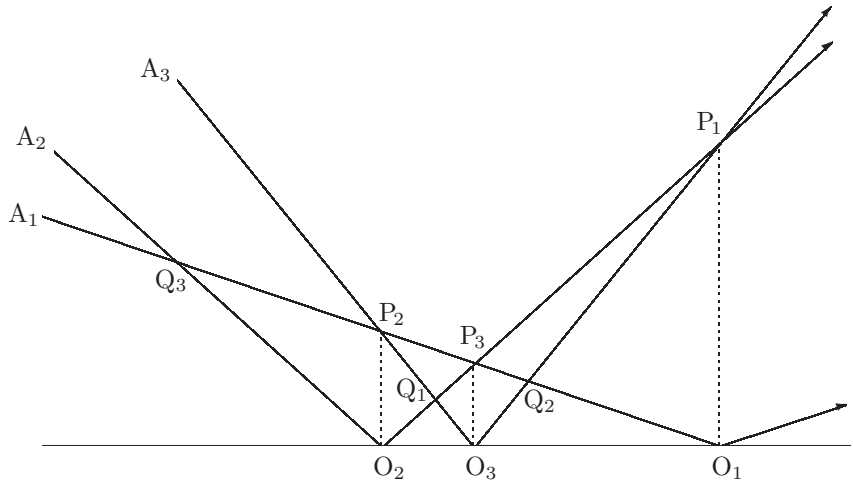
\begin{figure}[h]
\begin{center}
\begin{picture}(200,125)(-110,-5)
\put(-127,80){A$_2$}
\put(-30,0){\rotatebox[origin=l]{-42}{\line(-1,0){117}}}
\put(-33.6,-0.5){\rotatebox[origin=l]{42}{\vector(1,0){160}}}
\put(-94,98){A$_3$}
\put(-5.5,0){\rotatebox[origin=l]{-51.06}{\line(-1,0){125}}}
\put(-10,-1){\rotatebox[origin=l]{51.06}{\vector(1,0){150}}}
\put(-129,59.5){A$_1$}
\put(60,0){\rotatebox[origin=l]{-18.74}{\line(-1,0){190}}}
\put(58.5,0){\rotatebox[origin=l]{18.06}{\vector(1,0){35}}}
\drawline(-120,0)(95,0)
\multiput(60,0)(0,2){40}{\put(0,0){\line(0,1){0.6}}}
\multiput(-5.45455,0)(0,2){11}{\put(0,0){\line(0,1){0.6}}}
\multiput(-30,0)(0,2){15}{\put(0,0){\line(0,1){0.6}}}
\put(-30.5,33){P$_{\!2}$}\put(-9,27){P$_{\!3}$}\put(54,84){P$_{\!1}$}
\put(-32,-8){O$_2$}\put(-7.5,-8){O$_3$}\put(58,-8){O$_1$}
\put(-90,42){Q$_{3}$}\put(-26,12.9){Q$_{1}$}\put(8,10){Q$_{2}$}
\end{picture}
\end{center}
\caption{Scattering diagram for the RHS of (\ref{reika}).}
\label{RS1}
\end{figure}
\setlength\unitlength{1pt}

One can arrange the three particle world-lines so that 
the two particle scattering P$_1$, P$_2$, P$_3$ happen
exactly at the same instant as the boundary reflection O$_1$, O$_2$, O$_3$
of the other particle, respectively.
This is nontrivial.
For instance, suppose there were only particles 2 and 3.
They already determine the reflecting points O$_2$, O$_3$ and 
the intersection P$_1$ (and Q$_1$) 
and its projection O$_1$ onto the boundary.
Let P$_2$, P$_3$ be the points on the world-lines of particle 3 and 2 
whose projection are O$_2$ and O$_3$, respectively.
In order to be able to draw the world-line for the last particle 1,
the three points P$_2$, P$_3$ and O$_1$ must be collinear.
This is guaranteed by a special case of the Pappus theorem from the 4th century.

One can state it more symmetrically just by starting from 
P$_1$, P$_2$ and their projection O$_1$, O$_2$ onto the boundary.
Let P$_1'$, P$_2'$ be the mirror image of P$_1$, P$_2$
with respect to the boundary.
Then the three intersections
$\overline{{\rm P}_1{\rm O}_2} \cap \overline{{\rm O}_1{\rm P}_2}$,
$\overline{{\rm P}_1{\rm P}_2'} \cap \overline{{\rm P}_1'{\rm P}_2}$ and 
$\overline{{\rm O}_1{\rm P}_2'} \cap \overline{{\rm P}_1'{\rm O}_2}$
are collinear; in fact they are P$_3$, O$_3$ and the mirror image of P$_3$. 

Let us call the so arranged scattering diagram a {\em Pappus configuration}.
The reflection at O$_i$ with the simultaneous two particle scattering at P$_i$
will be referred to as a {\em special three particle event}\;($i=1,2,3$).
Up to a translation in the 
horizontal direction, 
a Pappus configuration is parameterized by three real numbers.
For instance one can specify it by 
the length of the segment $\overline{{\rm O}_1{\rm O}_2}$ and 
the (dual) reflection angles 
$\angle$P$_3$O$_2$O$_3$ and 
$\angle$P$_3$O$_1$O$_3$.
Set
\begin{equation}\label{silk0}
\begin{split}
u &= \angle{\rm P}_3{\rm O}_2{\rm O}_3,
\quad
w = \angle{\rm P}_2{\rm O}_3{\rm O}_2,
\quad
v = \angle{\rm P}_3{\rm O}_1{\rm O}_3,\\
\theta_1 &=\angle{\rm A}_2{\rm Q}_3{\rm A}_1,
\quad
\theta_2 = \angle{\rm A}_3{\rm P}_2{\rm A}_1,
\quad
\theta_3 = \angle{\rm A}_3{\rm Q}_1{\rm O}_2,\\
\theta_4 &= \angle{\rm A}_1{\rm P}_3{\rm O}_2,
\quad
\theta_5 = \angle{\rm A}_1{\rm Q}_2{\rm O}_3,
\quad
\theta_6 = \angle{\rm O}_2{\rm P}_1{\rm O}_3.
\end{split}
\end{equation}
Then it is elementary to see
\begin{align}
&\tan w = \tan u + \tan v,
\label{silk1}\\
&\theta_1 = u-v,\quad \theta_2 = w-v, \quad \theta_3 = u+w,\quad
\theta_4 = u+v, \quad \theta_5 = v+w, \quad \theta_6 = w-u.
\label{silk2}
\end{align}
We formally consider the infinitesimal angles hence replace (\ref{silk1}) by 
$w=u+v$.
In such a treatment, a Pappus configuration is labeled only by 
the two angles $u$ and $v$.
By a further substitution $u=\alpha_1+\alpha_2$ and $v = \alpha_2$,
(\ref{silk2}) becomes
\begin{align}\label{theta}
\theta_1 = \alpha_1, \;\;
\theta_2 = \alpha_1 + \alpha_2, \;\;
\theta_3 = 2\alpha_1 + 3\alpha_2, \;\;
\theta_4 = \alpha_1 + 2\alpha_2, \;\;
\theta_5 = \alpha_1 + 3\alpha_2, \;\;
\theta_6 = \alpha_2.
\end{align}
Regard the symbols $\alpha_1, \alpha_2$ formally as the simple roots of $G_2$.
They are transformed by the simple reflections $s_1, s_2$ of the Weyl group as
\begin{align*}
s_1(\alpha_1) = -\alpha_1, \quad s_1(\alpha_2) = \alpha_1+\alpha_2,
\quad
s_2(\alpha_1) = \alpha_1+3\alpha_2, \quad s_2(\alpha_2) = -\alpha_2.
\end{align*}
Thus we find 
\begin{align}\label{mirk}
\theta_k = s_{i_1} \cdots s_{i_{k-1}}(\alpha_{i_k}),\qquad
(i_1,i_2,i_3,i_4, i_5, i_6)=(1,2,1,2,1,2),
\end{align}
and $\{\theta_1,\ldots, \theta_6\}$ 
yields the set of the positive roots of $G_2$.

The RHS of the $G_2$ reflection equation (\ref{reika})
is obtained by attaching $R({\rm e}^{\theta_k})$ to the two particle scattering at 
Q$_i$ and $G({\rm e}^{\theta_k})$ to the special three particle event at 
P$_i$O$_i$ if it is the $k$th event starting from the left in Figure \ref{RS1}.
Setting ${\rm e}^u = x$ and ${\rm e}^v = y$, 
the assignment reads
\begin{align*}
R_{21}(x) : &\;\text{two particle scattering at Q$_3$},\\
G_{312}(xy) : &\;\text{special three particle event at P$_2$O$_2$}, \\
R_{32}(x^2y^3) : &\;\text{two particle scattering at Q$_1$},\\
G_{123}(xy^2) : &\;\text{special three particle event at P$_3$O$_3$}, \\
R_{13}(xy^3) : &\;\text{two particle scattering at Q$_2$},\\
G_{231}(y) : &\;\text{special three particle event at P$_1$O$_1$}.
\end{align*}
The indices for each operator 
correspond to the ordering of the relevant particles before the process.
For instance just before the special three particle event at P$_2$O$_2$,
the incoming particles are 3,1,2 from the top to the bottom, which is encoded in 
$G_{312}(xy)$.
The LHS of the $G_2$ reflection equation (\ref{reika}) 
represents the Pappus configuration in which 
the time ordering of the processes are reversed. 
See Figure \ref{LS1}.

\setlength\unitlength{0.5mm}
\begin{figure}[h]
\begin{center}
\begin{picture}(200,125)(-90,-5)
\put(-95,106){$2$}
\put(26,-0.5){\rotatebox[origin=l]{42}{\vector(1,0){117}}}
\put(29.7,0){\rotatebox[origin=l]{-42}{\line(-1,0){160}}}
\put(-94,118){$3$}
\put(1,-1){\rotatebox[origin=l]{51.06}{\vector(1,0){125}}}
\put(5.3,0.3){\rotatebox[origin=l]{-51.06}{\line(-1,0){150}}}
\put(-98,10){$1$}
\put(-61.5,0){\rotatebox[origin=l]{18.74}{\vector(1,0){190}}}
\put(-59.9,0){\rotatebox[origin=l]{-18.06}{\line(-1,0){35}}}
\drawline(120,0)(-95,0)
\multiput(-60,0)(0,2){40}{\put(0,0){\line(0,1){0.6}}}
\multiput(5.45455,0)(0,2){11}{\put(0,0){\line(0,1){0.6}}}
\multiput(30,0)(0,2){15}{\put(0,0){\line(0,1){0.6}}}
\end{picture}
\end{center}
\caption{Scattering diagram for the LHS of (\ref{reika}).}
\label{LS1}
\end{figure}
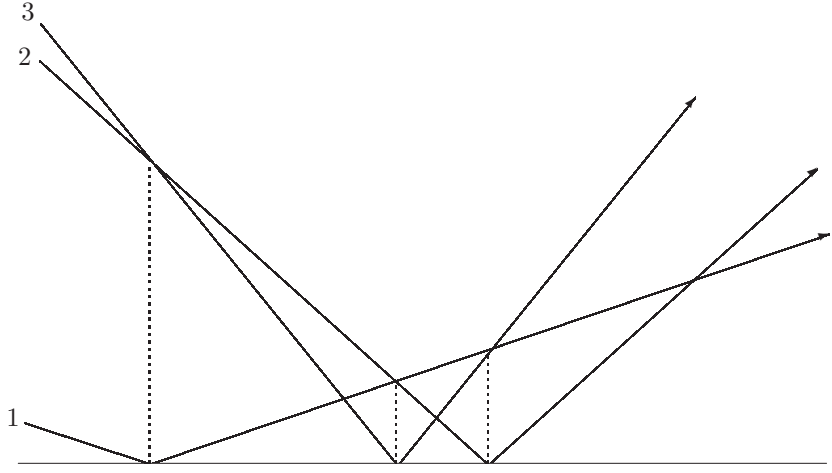
\setlength\unitlength{1pt}

\section{Quantized $G_2$ reflection equation}\label{sec:qgre}

\subsection{\mathversion{bold}$q$-bosons}
Let 
$F_q = \bigoplus_{m\ge 0}\C |m\rangle$ 
and $F_q^\ast = \bigoplus_{m \ge 0} \C\langle m |$ be 
the Fock space and its dual 
equipped with the inner product 
$\langle m | m'\rangle = (q^2)_m\delta_{m,m'}$.
We define the $q$-boson operators $\ap, \am, \ok$ on them by 
\begin{equation}\label{yum}
\begin{split}
&\ap |m\rangle = |m+1\rangle,\quad
\am |m\rangle = (1-q^{2m})|m-1\rangle,\quad
\ok |m\rangle = q^{m+\hf} |m\rangle,\\
&\langle m | \am = \langle m+1 |, \quad
\langle m | \ap = \langle m-1| (1-q^{2m}),\quad
\langle m | \ok = \langle m| q^{m+\hf}.
\end{split}
\end{equation}
They satisfy $(\langle m | X)|m'\rangle 
= \langle m | (X|m'\rangle)$.
Let $F_{q^3}, F^\ast_{q^3}$ 
and $\Ap, \Am, \OK$ denote 
the same objects with $q$ replaced by $q^3$. Namely,  
\begin{equation*}
\begin{split}
&\Ap |m\rangle = |m+1\rangle,\quad
\Am |m\rangle = (1-q^{6m})|m-1\rangle,\quad
\OK |m\rangle = q^{3m+\thf} |m\rangle,\\
&\langle m | \Am = \langle m+1 |, \quad
\langle m | \Ap = \langle m-1| (1-q^{6m}),\quad
\langle m | \OK = \langle m| q^{3m+\thf}.
\end{split}
\end{equation*}
The inner product in $F_{q^3}$ is given by 
$\langle m | m'\rangle = (q^6)_m\delta_{m,m'}$
differing from the $F_q$ case. 
However we write the base vectors as $\langle m |, |m\rangle$ 
either for 
$F^\ast_{q^3}, F_{q^3}$ or $F^\ast_{q}, F_{q}$
since their distinction will always be evident from the context.
Note the $q$-boson commutation relations
\begin{align}
&\ok \,\apm = q^{\pm 1}\apm \ok,\quad\quad\;
\apm \amp = 1 - q^{\mp 1}\ok^2,
\label{ngh1}\\
&\OK\, \Apm = q^{\pm 3}\Apm \OK,\quad
\Apm \Amp = 1 - q^{\mp 3}\OK^2.
\label{ngh2}
\end{align}
We will also use the number operator ${\bf h}$ defined by 
\begin{align}\label{syri}
{\bf h}|m\rangle = m | m\rangle,\qquad
\langle m | {\bf h}= \langle m| m
\end{align}
either for $F_q$ or $F_{q^3}$.
One may regard $\ok=q^{{\bf h}+\hf}$ and 
$\OK = q^{3{\bf h}+\thf}$.
The extra $1/2$ in the spectrum of $\log_q \ok$
is the {\em zero point energy},  which simplifies many forthcoming formulas.

\subsection{\mathversion{bold}$q$-boson valued $L$ matrix}\label{ss:L}
Set $V = \C v_0 \oplus \C v_1\simeq \C^2$.
This should not be confused with $\Vb$ in (\ref{RG}).
In fact they will be related as $\Vb= V^{\otimes n}$ later.
(See around (\ref{mikrup}).)
We introduce the $q$-boson valued 
$L$ matrix by
\begin{align}
&L(v_\alpha \otimes v_\beta \otimes |m\rangle)
= \sum_{\gamma,\delta \in\{0,1\}}v_\gamma\otimes v_\delta \otimes 
L^{\gamma, \delta}_{\alpha, \beta}|m\rangle,
\label{air}\\
&L= \begin{pmatrix}L^{\gamma,\delta}_{\alpha,\beta}\end{pmatrix} 
= \begin{pmatrix}
1 & 0 & 0 & 0 \\
0 & \OK & \Am & 0\\
0 & \Ap & -\OK & 0\\
0 & 0 & 0 & 1
\end{pmatrix} \in \mathrm{End}(V \otimes V \otimes F_{q^3}).
\label{Lop}
\end{align}
We attach a diagram to each component 
$L^{\gamma,\delta}_{\alpha,\beta} \in \mathrm{End}(F_{q^3})$ 
as follows\footnote{
The vertices here should be distinguished from 
those in Figure \ref{RS1} and \ref{LS1} since $V$ and $\Vb$ are different. }: 
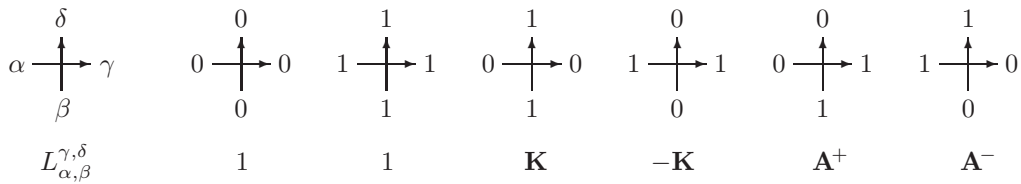
\begin{align*}
\begin{picture}(350.5,72)(6,35)
\put(12,80){
\put(-11,0){\vector(1,0){22}}\put(0,-10){\vector(0,1){20}}
}
\multiput(80,80)(55,0){6}{
\put(-11,0){\vector(1,0){22}}\put(0,-10){\vector(0,1){20}}
}
\put(-68,0){
\put(60,77){$\alpha$}\put(77.5,60){$\beta$}
\put(94.5,77){$\gamma$}\put(77.5,94){$\delta$}
}
\put(61,77){0}\put(77.5,60){0}\put(94,77){0}\put(77.5,94){0}
\put(55,0){
\put(61,77){1}\put(77.5,60){1}\put(94,77){1}\put(77.5,94){1}
}
\put(110,0){
\put(61,77){0}\put(77.5,60){1}\put(94,77){0}\put(77.5,94){1}
}
\put(165,0){
\put(61,77){1}\put(77.5,60){0}\put(94,77){1}\put(77.5,94){0}
}
\put(220,0){
\put(61,77){0}\put(77.5,60){1}\put(94,77){1}\put(77.5,94){0}
}
\put(275,0){
\put(61,77){1}\put(77.5,60){0}\put(94,77){0}\put(77.5,94){1}
}
\put(78,40){
\put(-74,0){$L^{\gamma,\delta}_{\alpha,\beta}$}
\put(0,0){1} \put(55,0){1} \put(109,0){$\OK$} 
\put(157,0){$-\OK$} \put(218,0){$\Ap$} \put(274,0){$\Am$}
}
\end{picture}
\end{align*}
The other configurations are to be understood as zero.
So $L$ may be regarded as defining a $q$-boson valued six vertex model
in which the latter relation of (\ref{ngh2}) plays the role of  
``free-fermion" condition.
See eq. $(10.16.5)|_{d=0}$ in \cite{Bax}.
Explicitly we have
\begin{align*}
L(v_0\otimes v_0 \otimes |m\rangle)
&= v_0 \otimes v_0 \otimes |m\rangle,
\quad
L(v_1\otimes v_1 \otimes |m\rangle)
= v_1 \otimes v_1 \otimes |m\rangle,
\\
L(v_0\otimes v_1 \otimes |m\rangle)
&= v_0 \otimes v_1 \otimes \OK|m\rangle
+ v_1 \otimes v_0 \otimes \Ap|m\rangle\\
&=  q^{3m+\thf}v_0 \otimes v_1 \otimes |m\rangle
+ v_1 \otimes v_0 \otimes |m+1\rangle,
\\
L(v_1\otimes v_0 \otimes |m\rangle)
&= v_0 \otimes v_1 \otimes \Am|m\rangle
- v_1 \otimes v_0 \otimes \OK|m\rangle\\
&= (1-q^{6m})v_0 \otimes v_1 \otimes |m-1\rangle
- q^{3m+\thf}v_1 \otimes v_0 \otimes |m\rangle.
\end{align*}
Note the obvious properties
\begin{align}
&L^{\gamma,\delta}_{\alpha,\beta}
=0
\quad \text{unless}\;\; \alpha+\beta=\gamma+\delta,
\label{mzsma1}\\
&{\bf h} L^{\gamma,\delta}_{\alpha,\beta}
=L^{\gamma,\delta}_{\alpha,\beta}
({\bf h}+\beta-\delta),
\label{mzsma2}
\end{align}
which will be referred to as {\em weight conservation}.
Up to conventional difference, the $L$ matrix (\ref{air}) 
appeared in \cite{BS}. See also \cite{KS, KP}.

\subsection{\mathversion{bold}$q$-boson valued $J$ matrix}\label{ss:J}
Besides the $L$, we need another $q$-boson valued matrix $J$
which encodes a characteristic feature of the 
$G_2$ scattering.
It is defined by 
\begin{align}
&J(v_\alpha \otimes v_\beta \otimes v_\gamma \otimes |m\rangle)
= \sum_{\lambda, \mu, \nu \in \{0,1\}}
v_{\lambda}\otimes v_{\mu}\otimes v_{\nu}\otimes 
J^{\lambda, \mu, \nu}_{\alpha, \beta, \gamma}|m\rangle,
\label{air1}\\
&J = \begin{pmatrix}
J^{\lambda, \mu, \nu}_{\alpha, \beta, \gamma}
\end{pmatrix}
\in \mathrm{End}(V \otimes V \otimes V \otimes F_q).
\label{air2}
\end{align}
Each component 
$J^{\lambda, \mu, \nu}_{\alpha, \beta, \gamma} \in \mathrm{End}(F_q)$
is depicted by a $90^{\circ}$-degrees rotated 
special three particle event\footnote{Note however again that 
the lines here carry $V$ 
whereas those in Figure \ref{RS1} and \ref{LS1} do $\Vb$.
The boundary line is omitted here.} 
\begin{align}\label{Jdef}
\begin{picture}(100,50)(-35,-20)
\put(-60,1){$J^{\lambda, \mu, \nu}_{\alpha, \beta, \gamma}=$}
\put(0,0){
\put(-13,21.5){$\mu$}\put(6,21){$\lambda$}\put(22,21){$\nu$}
\put(-9,-9){\vector(2,3){18}}\put(9,-9){\vector(-2,3){18}}
\multiput(0.1,4)(2,0){17}{\put(0,0){\line(1,0){0.4}}}
\put(34,3.9){\line(-2,-3){9}}\put(34,3.9){\vector(-2,3){9}}
\put(-13,-19){$\alpha$}\put(6,-20){$\beta$}\put(22,-19){$\gamma$}
}
\end{picture}
\end{align}
We choose the operator $J^{\lambda, \mu, \nu}_{\alpha, \beta, \gamma} 
\in \mathrm{End}(F_q)$
concretely as follows:
\begin{align}
\begin{picture}(400,70)(-60,-35)
\put(0,0){
\put(-13,21){$a$}\put(6,21){$a$}\put(22,20){0}
\put(-9,-9){\vector(2,3){18}}\put(9,-9){\vector(-2,3){18}}
\multiput(0.1,4)(2,0){17}{\put(0,0){\line(1,0){0.4}}}
\put(34,3.9){\line(-2,-3){9}}\put(34,3.9){\vector(-2,3){9}}
\put(-13,-19){$a$}\put(6,-19){$a$}\put(22,-20){0}
\put(5,-37){$\ap$}}
\put(80,0){
\put(-13,21){$a$}\put(6,21){$a$}\put(22,20){1}
\put(-9,-9){\vector(2,3){18}}\put(9,-9){\vector(-2,3){18}}
\multiput(0.1,4)(2,0){17}{\put(0,0){\line(1,0){0.4}}}
\put(34,3.9){\line(-2,-3){9}}\put(34,3.9){\vector(-2,3){9}}
\put(-13,-19){$a$}\put(6,-19){$a$}\put(22,-20){0}
\put(4,-37){$\ok$}}
\put(160,0){
\put(-13,21){$a$}\put(6,21){$a$}\put(22,20){0}
\put(-9,-9){\vector(2,3){18}}\put(9,-9){\vector(-2,3){18}}
\multiput(0.1,4)(2,0){17}{\put(0,0){\line(1,0){0.4}}}
\put(34,3.9){\line(-2,-3){9}}\put(34,3.9){\vector(-2,3){9}}
\put(-13,-19){$a$}\put(6,-19){$a$}\put(22,-20){1}
\put(0,-37){$-\ok$}}
\put(240,0){
\put(-13,21){$a$}\put(6,21){$a$}\put(22,20){1}
\put(-9,-9){\vector(2,3){18}}\put(9,-9){\vector(-2,3){18}}
\multiput(0.1,4)(2,0){17}{\put(0,0){\line(1,0){0.4}}}
\put(34,3.9){\line(-2,-3){9}}\put(34,3.9){\vector(-2,3){9}}
\put(-13,-19){$a$}\put(6,-19){$a$}\put(22,-20){1}
\put(5,-37){$\am$}}
\put(300,0){$(a=0,1)$}
\end{picture}
\end{align}
%
\begin{align}
\begin{picture}(400,70)(-60,-35)
\put(0,0){
\put(-13,20){1}\put(6,20){0}\put(22,20){0}
\put(-9,-9){\vector(2,3){18}}\put(9,-9){\vector(-2,3){18}}
\multiput(0.1,4)(2,0){17}{\put(0,0){\line(1,0){0.4}}}
\put(34,3.9){\line(-2,-3){9}}\put(34,3.9){\vector(-2,3){9}}
\put(-13,-20){0}\put(6,-20){1}\put(22,-20){0}
\put(0,-37){$u_1\ok\,\ap$}}
\put(80,0){
\put(-13,20){1}\put(6,20){0}\put(22,20){1}
\put(-9,-9){\vector(2,3){18}}\put(9,-9){\vector(-2,3){18}}
\multiput(0.1,4)(2,0){17}{\put(0,0){\line(1,0){0.4}}}
\put(34,3.9){\line(-2,-3){9}}\put(34,3.9){\vector(-2,3){9}}
\put(-13,-20){0}\put(6,-20){1}\put(22,-20){0}
\put(5,-37){$\ok^2$}}
\put(160,0){
\put(-13,20){1}\put(6,20){0}\put(22,20){0}
\put(-9,-9){\vector(2,3){18}}\put(9,-9){\vector(-2,3){18}}
\multiput(0.1,4)(2,0){17}{\put(0,0){\line(1,0){0.4}}}
\put(34,3.9){\line(-2,-3){9}}\put(34,3.9){\vector(-2,3){9}}
\put(-13,-20){0}\put(6,-20){1}\put(22,-20){1}
\put(-8,-37){$r^{-1}u_1u_3\sk$}}
\put(240,0){
\put(-13,20){1}\put(6,20){0}\put(22,20){1}
\put(-9,-9){\vector(2,3){18}}\put(9,-9){\vector(-2,3){18}}
\multiput(0.1,4)(2,0){17}{\put(0,0){\line(1,0){0.4}}}
\put(34,3.9){\line(-2,-3){9}}\put(34,3.9){\vector(-2,3){9}}
\put(-13,-20){0}\put(6,-20){1}\put(22,-20){1}
\put(0,-37){$u_3\ok\,\am$}}
\end{picture}
\end{align}
%
\begin{align}
\begin{picture}(400,70)(-60,-35)
\put(0,0){
\put(-13,20){0}\put(6,20){1}\put(22,20){0}
\put(-9,-9){\vector(2,3){18}}\put(9,-9){\vector(-2,3){18}}
\multiput(0.1,4)(2,0){17}{\put(0,0){\line(1,0){0.4}}}
\put(34,3.9){\line(-2,-3){9}}\put(34,3.9){\vector(-2,3){9}}
\put(-13,-20){1}\put(6,-20){0}\put(22,-20){0}
\put(-5,-37){$-u_2\, \ap \ok$}}
\put(80,0){
\put(-13,20){0}\put(6,20){1}\put(22,20){1}
\put(-9,-9){\vector(2,3){18}}\put(9,-9){\vector(-2,3){18}}
\multiput(0.1,4)(2,0){17}{\put(0,0){\line(1,0){0.4}}}
\put(34,3.9){\line(-2,-3){9}}\put(34,3.9){\vector(-2,3){9}}
\put(-13,-20){1}\put(6,-20){0}\put(22,-20){0}
\put(-8,-37){$r^{-1}u_2u_4\sk$}}
\put(160,0){
\put(-13,20){0}\put(6,20){1}\put(22,20){0}
\put(-9,-9){\vector(2,3){18}}\put(9,-9){\vector(-2,3){18}}
\multiput(0.1,4)(2,0){17}{\put(0,0){\line(1,0){0.4}}}
\put(34,3.9){\line(-2,-3){9}}\put(34,3.9){\vector(-2,3){9}}
\put(-13,-20){1}\put(6,-20){0}\put(22,-20){1}
\put(5,-37){$\ok^2$}}
\put(240,0){
\put(-13,20){0}\put(6,20){1}\put(22,20){1}
\put(-9,-9){\vector(2,3){18}}\put(9,-9){\vector(-2,3){18}}
\multiput(0.1,4)(2,0){17}{\put(0,0){\line(1,0){0.4}}}
\put(34,3.9){\line(-2,-3){9}}\put(34,3.9){\vector(-2,3){9}}
\put(-13,-20){1}\put(6,-20){0}\put(22,-20){1}
\put(-5,-37){$-u_4\am\ok$}}
\end{picture}
\end{align}
%
\begin{align}
\begin{picture}(400,70)(-60,-35)
\put(0,0){
\put(-13,20){0}\put(6,20){1}\put(22,20){0}
\put(-9,-9){\vector(2,3){18}}\put(9,-9){\vector(-2,3){18}}
\multiput(0.1,4)(2,0){17}{\put(0,0){\line(1,0){0.4}}}
\put(34,3.9){\line(-2,-3){9}}\put(34,3.9){\vector(-2,3){9}}
\put(-13,-20){0}\put(6,-20){1}\put(22,-20){0}
\put(2,-37){$(\ap)^2$}}
\put(80,0){
\put(-13,20){0}\put(6,20){1}\put(22,20){1}
\put(-9,-9){\vector(2,3){18}}\put(9,-9){\vector(-2,3){18}}
\multiput(0.1,4)(2,0){17}{\put(0,0){\line(1,0){0.4}}}
\put(34,3.9){\line(-2,-3){9}}\put(34,3.9){\vector(-2,3){9}}
\put(-13,-20){0}\put(6,-20){1}\put(22,-20){0}
\put(0,-37){$u_4\ok\,\ap$}}
\put(160,0){
\put(-13,20){0}\put(6,20){1}\put(22,20){0}
\put(-9,-9){\vector(2,3){18}}\put(9,-9){\vector(-2,3){18}}
\multiput(0.1,4)(2,0){17}{\put(0,0){\line(1,0){0.4}}}
\put(34,3.9){\line(-2,-3){9}}\put(34,3.9){\vector(-2,3){9}}
\put(-13,-20){0}\put(6,-20){1}\put(22,-20){1}
\put(-5,-37){$-u_3\, \ap \ok$}}
\put(240,0){
\put(-13,20){0}\put(6,20){1}\put(22,20){1}
\put(-9,-9){\vector(2,3){18}}\put(9,-9){\vector(-2,3){18}}
\multiput(0.1,4)(2,0){17}{\put(0,0){\line(1,0){0.4}}}
\put(34,3.9){\line(-2,-3){9}}\put(34,3.9){\vector(-2,3){9}}
\put(-13,-20){0}\put(6,-20){1}\put(22,-20){1}
\put(-8,-37){$r^{-1}u_3u_4\sk$}}
\end{picture}
\end{align}
%
\begin{align}
\begin{picture}(400,70)(-60,-35)
\put(0,0){
\put(-13,20){1}\put(6,20){0}\put(22,20){0}
\put(-9,-9){\vector(2,3){18}}\put(9,-9){\vector(-2,3){18}}
\multiput(0.1,4)(2,0){17}{\put(0,0){\line(1,0){0.4}}}
\put(34,3.9){\line(-2,-3){9}}\put(34,3.9){\vector(-2,3){9}}
\put(-13,-20){1}\put(6,-20){0}\put(22,-20){0}
\put(-8,-37){$r^{-1}u_1u_2\sk$}}
\put(80,0){
\put(-13,20){1}\put(6,20){0}\put(22,20){1}
\put(-9,-9){\vector(2,3){18}}\put(9,-9){\vector(-2,3){18}}
\multiput(0.1,4)(2,0){17}{\put(0,0){\line(1,0){0.4}}}
\put(34,3.9){\line(-2,-3){9}}\put(34,3.9){\vector(-2,3){9}}
\put(-13,-20){1}\put(6,-20){0}\put(22,-20){0}
\put(0,-37){$u_2\ok\,\am$}}
\put(160,0){
\put(-13,20){1}\put(6,20){0}\put(22,20){0}
\put(-9,-9){\vector(2,3){18}}\put(9,-9){\vector(-2,3){18}}
\multiput(0.1,4)(2,0){17}{\put(0,0){\line(1,0){0.4}}}
\put(34,3.9){\line(-2,-3){9}}\put(34,3.9){\vector(-2,3){9}}
\put(-13,-20){1}\put(6,-20){0}\put(22,-20){1}
\put(-5,-37){$-u_1\am\ok$}}
\put(240,0){
\put(-13,20){1}\put(6,20){0}\put(22,20){1}
\put(-9,-9){\vector(2,3){18}}\put(9,-9){\vector(-2,3){18}}
\multiput(0.1,4)(2,0){17}{\put(0,0){\line(1,0){0.4}}}
\put(34,3.9){\line(-2,-3){9}}\put(34,3.9){\vector(-2,3){9}}
\put(-13,-20){1}\put(6,-20){0}\put(22,-20){1}
\put(2,-37){$(\am)^2$}}
\end{picture}
\end{align}
Here $u_1, u_2, u_3, u_4$ are parameters satisfying
\begin{align}\label{rdef}
u_1u_2+u_3u_4=r:=q+q^{-1}.
\end{align} 
The operator $\sk \in \mathrm{End}(F_q)$ is defined by
\begin{align}\label{spd}
\sk = \am \ap- q^{-1}\ok^2 = 1-r \ok^2.
\end{align}
All the $J^{\lambda, \mu, \nu}_{\alpha, \beta, \gamma}$'s 
not contained in the above list is zero.
The weight conservation properties analogous to
(\ref{mzsma1}) and (\ref{mzsma2})  hold:.
\begin{align}
&J^{\lambda, \mu, \nu}_{\alpha, \beta, \gamma} = 0 
\;\; \text{unless}\;\; \alpha+\beta=\lambda + \mu,
\label{yumi1}\\
&{\bf h} J^{\lambda, \mu, \nu}_{\alpha, \beta, \gamma}
= J^{\lambda, \mu, \nu}_{\alpha, \beta, \gamma}
({\bf h}+1+\beta-\gamma-\mu-\nu).
\label{yumi2}
\end{align}
As an illustration we have
\begin{align*}
&J(v_1 \otimes v_0 \otimes v_0 \otimes |m \rangle)\\
&= v_1 \otimes v_0 \otimes v_0 \otimes  J^{100}_{100}|m \rangle
+ v_1 \otimes v_0 \otimes v_1 \otimes  J^{101}_{100}|m \rangle
\\
&+ v_0 \otimes v_1 \otimes v_0 \otimes  J^{010}_{100}|m \rangle
+ v_0 \otimes v_1 \otimes v_1 \otimes  J^{011}_{100}|m \rangle
\\
&= -u_2q^{m+\hf} v_1 \otimes v_0 \otimes v_0 \otimes  |m+1 \rangle
+r^{-1}u_2u_4(1-rq^{2m+1})v_1 \otimes v_0 \otimes v_1 \otimes |m \rangle\\
&+r^{-1}u_1u_2(1-rq^{2m+1})v_0 \otimes v_1 \otimes v_0 \otimes |m \rangle
+u_2q^{m-\hf}(1-q^{2m})v_0 \otimes v_1 \otimes v_1 \otimes |m-1 \rangle.
\end{align*}
The three particle diagram reduces to a direct product 
of two particle scattering and one particle boundary reflection
if the dotted line were absent.
Although it is {\em not} the case, 
the operator $J$ almost splits into such a product as
\begin{align}\label{gkkun}
J^{\lambda, \mu, \nu}_{\alpha, \beta, \gamma}= d 
{\mathcal L}^{\lambda, \mu}_{\alpha,\beta}\, K^{\nu}_{\gamma}
+c \theta(\alpha+\gamma=\mu+\nu=1)\mathrm{id}
\end{align}
for some constants $c,d$.
Here ${\mathcal L}$ denotes 
$(\ref{Lop})|_{\Apm \rightarrow \apm, \OK \rightarrow \ok}$
and $K^{\nu}_{\gamma}$ are the $q$-boson valued $K$ matrix 
introduced in \cite[eq.(9)]{KP}.

\subsection{\mathversion{bold}Quantized $G_2$ reflection equation}
\label{ss:qgre}

Given $L$ and $J$ in Section \ref{ss:L} and \ref{ss:J},
consider the $G_2$ reflection equation
$(\ref{reika})|_{R\rightarrow L, G \rightarrow J}$ 
that holds up to conjugation by an element 
$\Fm \in \mathrm{End}(F_{q^3}\otimes 
F_q \otimes F_{q^3}\otimes F_q \otimes F_{q^3}\otimes F_q)$:
\begin{align}\label{qre}
(L_{12}J_{132}L_{23}J_{213}L_{31}J_{321}) \circ \Fm
=\Fm \circ (J_{231}L_{13}J_{123}L_{32}J_{312}L_{21}).
\end{align}
This is an equality of linear operators on
$\overset{1}{V}\otimes \overset{2}{V}\otimes \overset{3}{V} 
\otimes \overset{4}{F}_{q^3} 
\otimes \overset{5}{F}_{q} 
\otimes \overset{6}{F}_{q^3} 
\otimes \overset{7}{F}_{q}
\otimes \overset{8}{F}_{q^3} 
\otimes \overset{9}{F}_{q}$,
where the superscripts are just temporal labels for explanation. 
If they are all exhibited (\ref{qre}) reads as 
\begin{align}\label{hrk}
L_{124}J_{1325}L_{236}J_{2137}L_{318}J_{3219} 
\Fm_{456789}
=
\Fm_{456789}
J_{2319}L_{138}J_{1237}L_{326}J_{3125}L_{214}.
\end{align}
We fix the normalization of $\Fm$ by 
\begin{align}\label{tsgmi}
\Fm (|0 \rangle \otimes |0 \rangle \otimes
|0 \rangle \otimes|0 \rangle\otimes|0 \rangle\otimes|0 \rangle) = 
|0 \rangle \otimes |0 \rangle \otimes
|0 \rangle \otimes|0 \rangle
\otimes|0 \rangle\otimes|0 \rangle.
\end{align}

To explain the notation in (\ref{qre}) and 
(\ref{hrk}), write $L$ (\ref{air}) and $J$ (\ref{air1}) as
$L = \sum \mathcal{L}^{(1)}_l \otimes \mathcal{L}^{(2)}_l 
\otimes \mathcal{L}^{(3)}_l$
and 
$J = \sum \mathcal{J}^{(1)}_l \otimes \mathcal{J}^{(2)}_l 
\otimes \mathcal{J}^{(3)}_l \otimes \mathcal{J}^{(4)}_l $ 
similarly to (\ref{ruby}), where $\sum$ means $\sum_l$. 
Then
\begin{align*}
L_{ij4} &= \sum \mathcal{L}_l^{(i)} \otimes \mathcal{L}_l^{(j)} \otimes
\ichi \otimes \mathcal{L}_l^{(3)} 
\otimes \ichi
\otimes \ichi \otimes \ichi 
\otimes \ichi \otimes \ichi\quad ((i,j)=(1,2),(2,1)),\\
L_{ij6} &= \sum \ichi \otimes 
\mathcal{L}_l^{(i-1)} \otimes \mathcal{L}_l^{(j-1)} 
\otimes \ichi 
\otimes \ichi
 \otimes \mathcal{L}_l^{(3)} \otimes \ichi 
\otimes \ichi \otimes \ichi\quad ((i,j)=(2,3),(3,2)),\\
L_{ij8} &= \sum  
\mathcal{L}_l^{(i')} \otimes \ichi \otimes \mathcal{L}_l^{(j')} 
\otimes \ichi 
 \otimes \ichi \otimes \ichi
\otimes \ichi  \otimes \mathcal{L}_l^{(3)}\otimes \ichi 
\quad ((i,j)=(1,3),(3,1), 1'=1, 3'=2),\\
J_{ijk5} & = \sum 
\mathcal{J}^{(i)}_l \otimes \mathcal{J}^{(j)}_l \otimes \mathcal{J}^{(k)}_l
\otimes \ichi \otimes \mathcal{J}^{(4)}_l 
\otimes \ichi\otimes \ichi\otimes \ichi\otimes \ichi \quad
(\{i,j,k\}=\{1,2,3\}),\\
J_{ijk7} & = \sum 
\mathcal{J}^{(i)}_l \otimes \mathcal{J}^{(j)}_l \otimes \mathcal{J}^{(k)}_l
\otimes \ichi \otimes \ichi \otimes \ichi \otimes \mathcal{J}^{(4)}_l 
\otimes \ichi\otimes \ichi \quad
(\{i,j,k\}=\{1,2,3\}),\\
J_{ijk9} & = \sum 
\mathcal{J}^{(i)}_l \otimes \mathcal{J}^{(j)}_l \otimes \mathcal{J}^{(k)}_l
\otimes \ichi \otimes \ichi \otimes \ichi \otimes
\ichi \otimes \ichi \otimes \mathcal{J}^{(4)}_l \quad
(\{i,j,k\}=\{1,2,3\}).
\end{align*}
Practically, one can realize these operators 
from (\ref{air}) and (\ref{air1})
by putting $L^{\gamma, \delta}_{\alpha, \beta}$ and 
$J^{\lambda, \mu, \nu}_{\alpha, \beta, \gamma}$ 
at appropriate tensor components with a suitable permutations of the indices 
$\alpha, \beta, \ldots$.
The equation (\ref{qre}) or equivalently (\ref{hrk}) is  
a $q$-boson valued $G_2$ reflection equation without a spectral parameter
up to conjugation.
We call them the 
{\em quantized $G_2$ reflection equation} in analogy with the 
quantized reflection equation proposed in \cite{KP} for $C_2$.
It is depicted as follows.

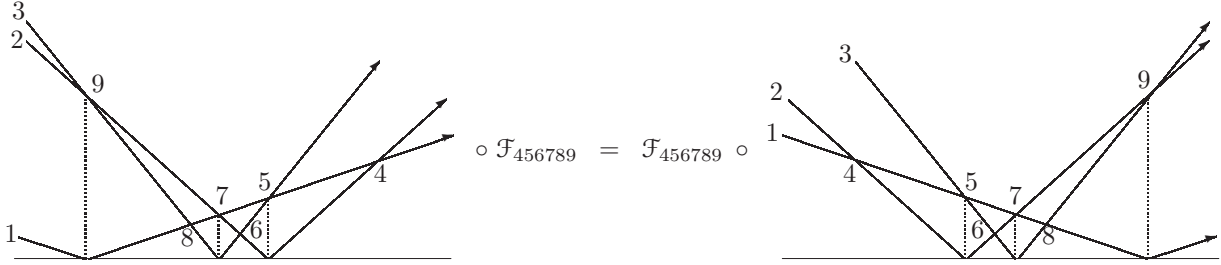
\setlength\unitlength{0.27mm}
\begin{align*} 
\begin{picture}(320,140)(50,-15)
\put(-97,104){$2$}
\put(22,-1){\rotatebox[origin=l]{42}{\vector(1,0){117}}}
\put(29.7,0){\rotatebox[origin=l]{-42}{\line(-1,0){160}}}
\put(-96,118){$3$}
\put(-4,-2){\rotatebox[origin=l]{51.06}{\vector(1,0){125}}}
\put(5.,0.3){\rotatebox[origin=l]{-51.06}{\line(-1,0){150}}}
\put(-100,8){$1$}
\put(-63,-0.3){\rotatebox[origin=l]{18.74}{\vector(1,0){190}}}
\put(-59.9,0){\rotatebox[origin=l]{-18.06}{\line(-1,0){35}}}
\drawline(120,0)(-95,0)
\multiput(-60,0)(0,2){40}{\put(0,0){\line(0,1){0.6}}}
\multiput(5.45455,0)(0,2){11}{\put(0,0){\line(0,1){0.6}}}
\multiput(30,0)(0,2){15}{\put(0,0){\line(0,1){0.6}}}
\put(25,34){5}\put(4,27){7}\put(-57,82){9}
\put(82,37.5){4}\put(21,10){6}\put(-13,6){8}
\put(132,50){$\circ\; \Fm_{456789}$}
\put(403,0){
\put(-210,50){$=\;\; \Fm_{456789}\; \,\circ$}
\put(-126,78){2}
\put(-30,0){\rotatebox[origin=l]{-42}{\line(-1,0){117}}}
\put(-37.5,-1){\rotatebox[origin=l]{42}{\vector(1,0){160}}}
\put(-92,97){3}
\put(-5.5,0){\rotatebox[origin=l]{-51.06}{\line(-1,0){125}}}
\put(-14,-2){\rotatebox[origin=l]{51.06}{\vector(1,0){150}}}
\put(-128,58){1}
\put(60,0){\rotatebox[origin=l]{-18.74}{\line(-1,0){190}}}
\put(56.5,0){\rotatebox[origin=l]{18.06}{\vector(1,0){35}}}
\drawline(-120,0)(95,0)
\multiput(60,0)(0,2){40}{\put(0,0){\line(0,1){0.6}}}
\multiput(-5.45455,0)(0,2){11}{\put(0,0){\line(0,1){0.6}}}
\multiput(-30,0)(0,2){15}{\put(0,0){\line(0,1){0.6}}}
\put(-30,34){5}\put(-8.5,27){7}\put(55,84){9}
\put(-90,38){4}\put(-27,10.3){6}\put(8,7){8}
}
\end{picture}
\end{align*}
\setlength\unitlength{1pt}
Here the indices 1,2,3 label the reflecting lines while 
4,5,6,7,8,9 are attached to the scattering/reflection events.
The latter group of indices are associated with 
the Fock spaces, and the $q$-bosons are 
acting on them in the direction perpendicular to this planar diagram.  
If one introduces such $q$-boson arrows 
going from the back to the front of the diagram, 
the operator $\Fm_{456789}$ in the LHS (resp.~RHS)
corresponds to a vertex where the six arrows going toward (resp. coming from)
4,5,6,7,8,9 intersect.
In Section \ref{ss:cc} we will take the concatenation of (\ref{qre}) for $n$ times. 
It corresponds to a 3D diagram involving the $n$ {\em layers} 
of the Pappus configurations
depicted in the above.  

The component of (\ref{hrk}) corresponding to the transition 
$v_i \otimes v_j \otimes v_k \mapsto
v_a \otimes v_b \otimes v_c$ in 
$\overset{1}{V}\otimes \overset{2}{V}\otimes \overset{3}{V}$
is given by 
\begin{equation}\label{minami}
\begin{split}
&\left(\sum
L^{a,b}_{\alpha_1, \alpha_2} \otimes 
J^{\alpha_1, c, \alpha_2}_{\beta_1, \beta_3, \beta_2} \otimes
L^{\beta_2, \beta_3}_{\gamma_1, \gamma_2} \otimes 
J^{\gamma_1, \beta_1, \gamma_2}_{\lambda_2, \mu_1, \mu_2} \otimes
L^{\mu_2, \mu_1}_{\lambda_3, \lambda_1} \otimes 
J^{\lambda_3, \lambda_2, \lambda_1}_{k,j,i}\right) \mathscr{F}\\
&= \mathscr{F}\left(\sum
L^{\alpha_2, \alpha_1}_{j,i} \otimes
J^{\beta_3, \beta_1, \beta_2}_{k,\alpha_1, \alpha_2}\otimes 
L^{\gamma_2, \gamma_1}_{\beta_3, \beta_2} \otimes 
J^{\mu_1, \lambda_2, \mu_2}_{\beta_1, \gamma_1, \gamma_2} \otimes 
L^{\lambda_1, \lambda_3}_{\mu_1, \mu_2} \otimes 
J^{b,c,a}_{\lambda_2, \lambda_3, \lambda_1}
\right),
\end{split}
\end{equation}
with the sums taken over 
$\alpha_1, \alpha_2, \beta_1,\beta_2,\beta_3,
\gamma_1, \gamma_2, \lambda_1, \lambda_2, \lambda_3,
\mu_1, \mu_2 \in \{0,1\}$.
The summands correspond to various diagrams  
with the external edges specified as

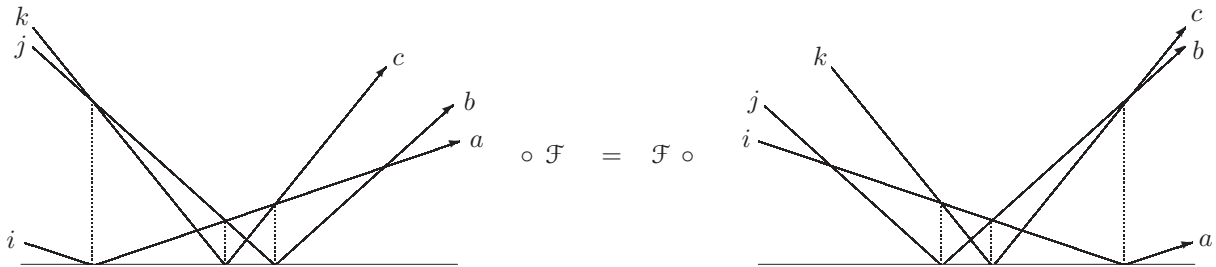
\setlength\unitlength{0.27mm}
\begin{align*} 
\begin{picture}(320,120)(35,2)
\put(-98,104){$j$}
\put(22,-1){\rotatebox[origin=l]{42}{\vector(1,0){117}}}
\put(29.7,0){\rotatebox[origin=l]{-42}{\line(-1,0){160}}}
\put(-98,118){$k$}
\put(-4,-2){\rotatebox[origin=l]{51.06}{\vector(1,0){125}}}
\put(5.,0.3){\rotatebox[origin=l]{-51.06}{\line(-1,0){150}}}
\put(-102,8){$i$}
\put(-63,-0.3){\rotatebox[origin=l]{18.74}{\vector(1,0){190}}}
\put(-59.9,0){\rotatebox[origin=l]{-18.06}{\line(-1,0){35}}}
\drawline(120,0)(-95,0)
\multiput(-60,0)(0,2){40}{\put(0,0){\line(0,1){0.6}}}
\multiput(5.45455,0)(0,2){11}{\put(0,0){\line(0,1){0.6}}}
\multiput(30,0)(0,2){15}{\put(0,0){\line(0,1){0.6}}}
\put(126,58){$a$}\put(123,76){$b$}\put(88,98){$c$}
\put(151,50){$\circ\; \,\Fm$}
\put(388,0){
\put(-197,50){$=\;\; \;\Fm\; \,\circ$}
\put(-125,77){$j$}
\put(-30,0){\rotatebox[origin=l]{-42}{\line(-1,0){117}}}
\put(-37.5,-1){\rotatebox[origin=l]{42}{\vector(1,0){160}}}
\put(-93,98){$k$}
\put(-5.5,0){\rotatebox[origin=l]{-51.06}{\line(-1,0){125}}}
\put(-14,-2){\rotatebox[origin=l]{51.06}{\vector(1,0){150}}}
\put(-128,58){$i$}
\put(60,0){\rotatebox[origin=l]{-18.74}{\line(-1,0){190}}}
\put(56.5,0){\rotatebox[origin=l]{18.06}{\vector(1,0){35}}}
\drawline(-120,0)(95,0)
\multiput(60,0)(0,2){40}{\put(0,0){\line(0,1){0.6}}}
\multiput(-5.45455,0)(0,2){11}{\put(0,0){\line(0,1){0.6}}}
\multiput(-30,0)(0,2){15}{\put(0,0){\line(0,1){0.6}}}
\put(97,9){$a$}\put(94,101){$b$}\put(93,120){$c$}
}
\end{picture}
\end{align*}
\setlength\unitlength{1pt}

Let us illustrate the case $(a,b,c,i,j,k)=(1,1,1,1,0,0)$.
The relevant diagrams are given as follows:

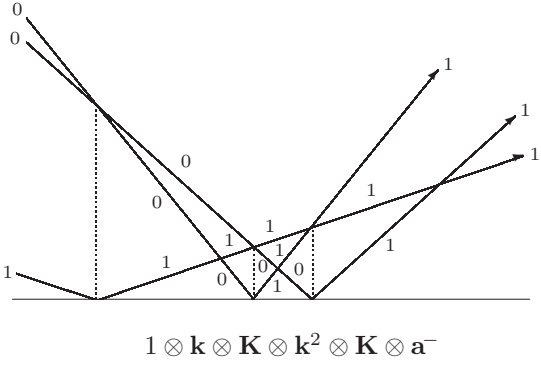
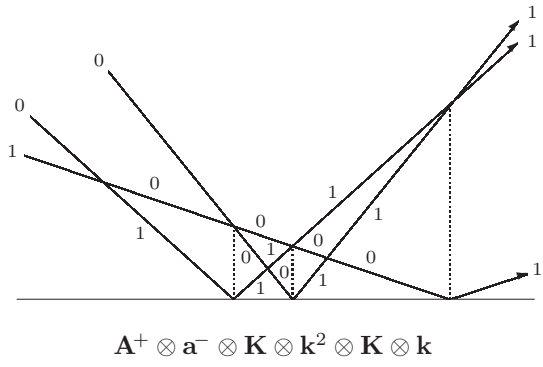
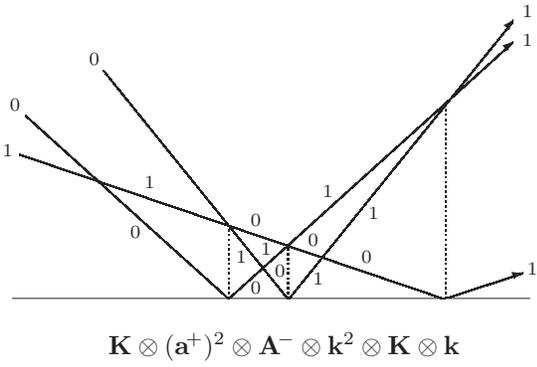
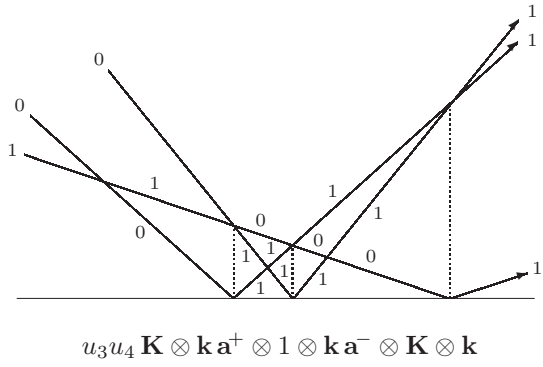
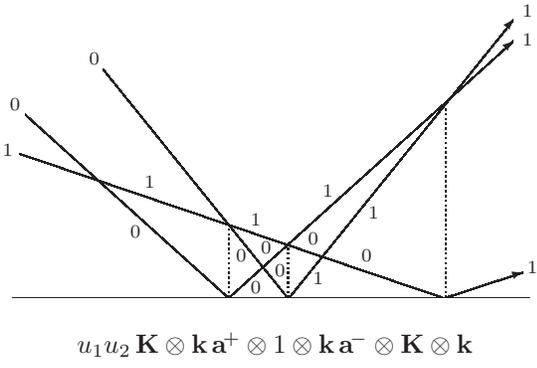
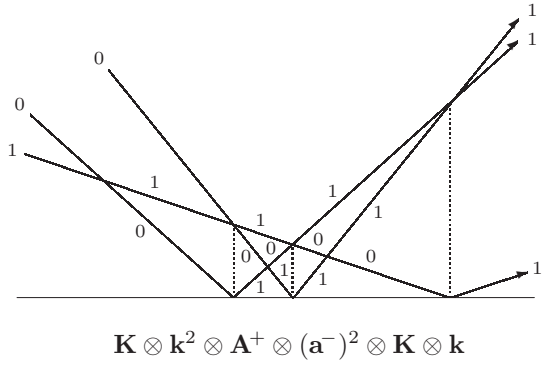
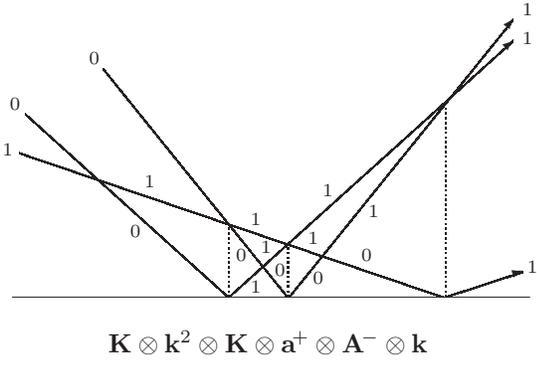
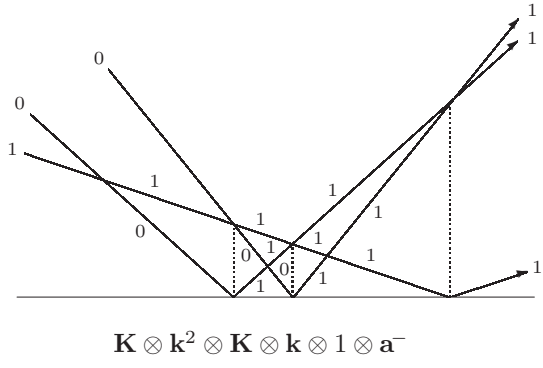
\setlength\unitlength{0.32mm}
\begin{align*} 
\begin{picture}(500,145)(-105,-20)
\put(-95,118){$\scriptstyle{0}$}
\put(-96,106){$\scriptstyle{0}$}
\put(-99,9){$\scriptstyle{1}$}
\put(84,95){$\scriptstyle{1}$}
\put(116,76){$\scriptstyle{1}$}
\put(120,58){$\scriptstyle{1}$}
\multiput(-60,0)(0,2){40}{\put(0,0){\line(0,1){0.6}}}
\multiput(5.45455,0)(0,2){11}{\put(0,0){\line(0,1){0.6}}}
\multiput(30,0)(0,2){15}{\put(0,0){\line(0,1){0.6}}}
\put(22,-1){\rotatebox[origin=l]{42}{\vector(1,0){114}}}
\put(29.7,0){\rotatebox[origin=l]{-42}{\line(-1,0){160}}}
\put(-4,-2){\rotatebox[origin=l]{51.06}{\vector(1,0){123}}}
\put(5.,0.3){\rotatebox[origin=l]{-51.06}{\line(-1,0){150}}}
\put(-63,-0.3){\rotatebox[origin=l]{18.74}{\vector(1,0){187}}}
\put(-59.9,0){\rotatebox[origin=l]{-18.06}{\line(-1,0){35}}}
\drawline(120,0)(-95,0)
\put(52,43){$\scriptstyle{1}$}
\put(10,28){$\scriptstyle{1}$}
\put(7,11){$\scriptstyle{0}$}
\put(14,18){$\scriptstyle{1}$}
\put(13,3){$\scriptstyle{1}$}
\put(-25,55){$\scriptstyle{0}$}
\put(-37,38){$\scriptstyle{0}$}
\put(-33,13){$\scriptstyle{1}$}
\put(60,20){$\scriptstyle{1}$}
\put(22,10){$\scriptstyle{0}$}
\put(-10,6){$\scriptstyle{0}$}
\put(-7,22){$\scriptstyle{1}$}
\put(-40,-23){$\ichi \otimes \ok \otimes \OK \otimes \ok^2 
\otimes \OK \otimes \am$}
%
\put(280,0){
\put(-88,97){$\scriptstyle{0}$}
\put(-121,78){$\scriptstyle{0}$}
\put(-124,59){$\scriptstyle{1}$}
\put(92,117){$\scriptstyle{1}$}
\put(92,105){$\scriptstyle{1}$}
\put(94,10){$\scriptstyle{1}$}
\put(9,42){$\scriptstyle{1}$}
\put(28,33){$\scriptstyle{1}$}
\put(-30,0){\rotatebox[origin=l]{-42}{\line(-1,0){114}}}
\put(-37.5,-1){\rotatebox[origin=l]{42}{\vector(1,0){159}}}
\put(-5.5,0){\rotatebox[origin=l]{-51.06}{\line(-1,0){122}}}
\put(-14,-2){\rotatebox[origin=l]{51.06}{\vector(1,0){149}}}
\put(60,0){\rotatebox[origin=l]{-18.74}{\line(-1,0){187}}}
\put(56.5,0){\rotatebox[origin=l]{18.06}{\vector(1,0){34}}}
\drawline(-120,0)(95,0)
\multiput(60,0)(0,2){40}{\put(0,0){\line(0,1){0.6}}}
\multiput(-5.45455,0)(0,2){11}{\put(0,0){\line(0,1){0.6}}}
\multiput(-30,0)(0,2){15}{\put(0,0){\line(0,1){0.6}}}
\put(-65,45.6){$\scriptstyle{0}$}
\put(-71,25){$\scriptstyle{1}$}
\put(-21,30){$\scriptstyle{0}$}
\put(-27,15){$\scriptstyle{0}$}
\put(-21,2){$\scriptstyle{1}$}
\put(-16.6,18.3){$\scriptstyle{1}$}
\put(-11,9){$\scriptstyle{0}$}
\put(3,22){$\scriptstyle{0}$}
\put(5,5.5){$\scriptstyle{1}$}
\put(25,15){$\scriptstyle{0}$}
\put(-80,-23){$\Ap \otimes \am \otimes \OK 
\otimes \ok^2 \otimes \OK \otimes \ok$}
}
\end{picture}
\end{align*}
\begin{align*} 
\begin{picture}(500,145)(-105,-20)
\put(25,0){
\put(-88,97){$\scriptstyle{0}$}
\put(-121,78){$\scriptstyle{0}$}
\put(-124,59){$\scriptstyle{1}$}
\put(92,117){$\scriptstyle{1}$}
\put(92,105){$\scriptstyle{1}$}
\put(94,10){$\scriptstyle{1}$}
\put(9,42){$\scriptstyle{1}$}
\put(28,33){$\scriptstyle{1}$}
\put(-30,0){\rotatebox[origin=l]{-42}{\line(-1,0){114}}}
\put(-37.5,-1){\rotatebox[origin=l]{42}{\vector(1,0){159}}}
\put(-5.5,0){\rotatebox[origin=l]{-51.06}{\line(-1,0){122}}}
\put(-14,-2){\rotatebox[origin=l]{51.06}{\vector(1,0){149}}}
\put(60,0){\rotatebox[origin=l]{-18.74}{\line(-1,0){187}}}
\put(56.5,0){\rotatebox[origin=l]{18.06}{\vector(1,0){34}}}
\drawline(-120,0)(95,0)
\multiput(60,0)(0,2){40}{\put(0,0){\line(0,1){0.6}}}
\multiput(-5.45455,0)(0,2){11}{\put(0,0){\line(0,1){0.6}}}
\multiput(-30,0)(0,2){15}{\put(0,0){\line(0,1){0.6}}}
\put(-65,45.6){$\scriptstyle{1}$}
\put(-71,25){$\scriptstyle{0}$}
\put(-21,30){$\scriptstyle{0}$}
\put(-27,15){$\scriptstyle{1}$}
\put(-21,2){$\scriptstyle{0}$}
\put(-16.6,18.3){$\scriptstyle{1}$}
\put(-11,9){$\scriptstyle{0}$}
\put(3,22){$\scriptstyle{0}$}
\put(5,5.5){$\scriptstyle{1}$}
\put(25,15){$\scriptstyle{0}$}
\put(-80,-23){$\OK \otimes (\ap)^2 \otimes \Am 
\otimes \ok^2 \otimes \OK \otimes \ok$}
}
\put(280,0){
\put(-88,97){$\scriptstyle{0}$}
\put(-121,78){$\scriptstyle{0}$}
\put(-124,59){$\scriptstyle{1}$}
\put(92,117){$\scriptstyle{1}$}
\put(92,105){$\scriptstyle{1}$}
\put(94,10){$\scriptstyle{1}$}
\put(9,42){$\scriptstyle{1}$}
\put(28,33){$\scriptstyle{1}$}
\put(-30,0){\rotatebox[origin=l]{-42}{\line(-1,0){114}}}
\put(-37.5,-1){\rotatebox[origin=l]{42}{\vector(1,0){159}}}
\put(-5.5,0){\rotatebox[origin=l]{-51.06}{\line(-1,0){122}}}
\put(-14,-2){\rotatebox[origin=l]{51.06}{\vector(1,0){149}}}
\put(60,0){\rotatebox[origin=l]{-18.74}{\line(-1,0){187}}}
\put(56.5,0){\rotatebox[origin=l]{18.06}{\vector(1,0){34}}}
\drawline(-120,0)(95,0)
\multiput(60,0)(0,2){40}{\put(0,0){\line(0,1){0.6}}}
\multiput(-5.45455,0)(0,2){11}{\put(0,0){\line(0,1){0.6}}}
\multiput(-30,0)(0,2){15}{\put(0,0){\line(0,1){0.6}}}
\put(-65,45.6){$\scriptstyle{1}$}
\put(-71,25){$\scriptstyle{0}$}
\put(-21,30){$\scriptstyle{0}$}
\put(-27,15){$\scriptstyle{1}$}
\put(-21,2){$\scriptstyle{1}$}
\put(-16.6,18.3){$\scriptstyle{1}$}
\put(-11,9){$\scriptstyle{1}$}
\put(3,22){$\scriptstyle{0}$}
\put(5,5.5){$\scriptstyle{1}$}
\put(25,15){$\scriptstyle{0}$}
\put(-93,-23){$u_3u_4\,\OK \otimes \ok\,\ap \otimes \ichi 
\otimes \ok\,\am \otimes \OK \otimes \ok$}
}
\end{picture}
\end{align*}
\begin{align*} 
\begin{picture}(500,145)(-105,-20)
\put(25,0){
\put(-88,97){$\scriptstyle{0}$}
\put(-121,78){$\scriptstyle{0}$}
\put(-124,59){$\scriptstyle{1}$}
\put(92,117){$\scriptstyle{1}$}
\put(92,105){$\scriptstyle{1}$}
\put(94,10){$\scriptstyle{1}$}
\put(9,42){$\scriptstyle{1}$}
\put(28,33){$\scriptstyle{1}$}
\put(-30,0){\rotatebox[origin=l]{-42}{\line(-1,0){114}}}
\put(-37.5,-1){\rotatebox[origin=l]{42}{\vector(1,0){159}}}
\put(-5.5,0){\rotatebox[origin=l]{-51.06}{\line(-1,0){122}}}
\put(-14,-2){\rotatebox[origin=l]{51.06}{\vector(1,0){149}}}
\put(60,0){\rotatebox[origin=l]{-18.74}{\line(-1,0){187}}}
\put(56.5,0){\rotatebox[origin=l]{18.06}{\vector(1,0){34}}}
\drawline(-120,0)(95,0)
\multiput(60,0)(0,2){40}{\put(0,0){\line(0,1){0.6}}}
\multiput(-5.45455,0)(0,2){11}{\put(0,0){\line(0,1){0.6}}}
\multiput(-30,0)(0,2){15}{\put(0,0){\line(0,1){0.6}}}
\put(-65,45.6){$\scriptstyle{1}$}
\put(-71,25){$\scriptstyle{0}$}
\put(-21,30){$\scriptstyle{1}$}
\put(-27,15){$\scriptstyle{0}$}
\put(-21,2){$\scriptstyle{0}$}
\put(-16.6,18.3){$\scriptstyle{0}$}
\put(-11,9){$\scriptstyle{0}$}
\put(3,22){$\scriptstyle{0}$}
\put(5,5.5){$\scriptstyle{1}$}
\put(25,15){$\scriptstyle{0}$}
\put(-93,-23){$u_1u_2\, \OK \otimes \ok\, \ap \otimes \ichi
\otimes \ok\, \am \otimes \OK \otimes \ok$}
}
\put(280,0){
\put(-88,97){$\scriptstyle{0}$}
\put(-121,78){$\scriptstyle{0}$}
\put(-124,59){$\scriptstyle{1}$}
\put(92,117){$\scriptstyle{1}$}
\put(92,105){$\scriptstyle{1}$}
\put(94,10){$\scriptstyle{1}$}
\put(9,42){$\scriptstyle{1}$}
\put(28,33){$\scriptstyle{1}$}
\put(-30,0){\rotatebox[origin=l]{-42}{\line(-1,0){114}}}
\put(-37.5,-1){\rotatebox[origin=l]{42}{\vector(1,0){159}}}
\put(-5.5,0){\rotatebox[origin=l]{-51.06}{\line(-1,0){122}}}
\put(-14,-2){\rotatebox[origin=l]{51.06}{\vector(1,0){149}}}
\put(60,0){\rotatebox[origin=l]{-18.74}{\line(-1,0){187}}}
\put(56.5,0){\rotatebox[origin=l]{18.06}{\vector(1,0){34}}}
\drawline(-120,0)(95,0)
\multiput(60,0)(0,2){40}{\put(0,0){\line(0,1){0.6}}}
\multiput(-5.45455,0)(0,2){11}{\put(0,0){\line(0,1){0.6}}}
\multiput(-30,0)(0,2){15}{\put(0,0){\line(0,1){0.6}}}
\put(-65,45.6){$\scriptstyle{1}$}
\put(-71,25){$\scriptstyle{0}$}
\put(-21,30){$\scriptstyle{1}$}
\put(-27,15){$\scriptstyle{0}$}
\put(-21,2){$\scriptstyle{1}$}
\put(-16.6,18.3){$\scriptstyle{0}$}
\put(-11,9){$\scriptstyle{1}$}
\put(3,22){$\scriptstyle{0}$}
\put(5,5.5){$\scriptstyle{1}$}
\put(25,15){$\scriptstyle{0}$}
\put(-80,-23){$\OK \otimes \ok^2 \otimes \Ap 
\otimes (\am)^2 \otimes \OK \otimes \ok$}
}
\end{picture}
\end{align*}
\begin{align*} 
\begin{picture}(500,145)(-105,-20)
\put(25,0){
\put(-88,97){$\scriptstyle{0}$}
\put(-121,78){$\scriptstyle{0}$}
\put(-124,59){$\scriptstyle{1}$}
\put(92,117){$\scriptstyle{1}$}
\put(92,105){$\scriptstyle{1}$}
\put(94,10){$\scriptstyle{1}$}
\put(9,42){$\scriptstyle{1}$}
\put(28,33){$\scriptstyle{1}$}
\put(-30,0){\rotatebox[origin=l]{-42}{\line(-1,0){114}}}
\put(-37.5,-1){\rotatebox[origin=l]{42}{\vector(1,0){159}}}
\put(-5.5,0){\rotatebox[origin=l]{-51.06}{\line(-1,0){122}}}
\put(-14,-2){\rotatebox[origin=l]{51.06}{\vector(1,0){149}}}
\put(60,0){\rotatebox[origin=l]{-18.74}{\line(-1,0){187}}}
\put(56.5,0){\rotatebox[origin=l]{18.06}{\vector(1,0){34}}}
\drawline(-120,0)(95,0)
\multiput(60,0)(0,2){40}{\put(0,0){\line(0,1){0.6}}}
\multiput(-5.45455,0)(0,2){11}{\put(0,0){\line(0,1){0.6}}}
\multiput(-30,0)(0,2){15}{\put(0,0){\line(0,1){0.6}}}
\put(-65,45.6){$\scriptstyle{1}$}
\put(-71,25){$\scriptstyle{0}$}
\put(-21,30){$\scriptstyle{1}$}
\put(-27,15){$\scriptstyle{0}$}
\put(-21,2){$\scriptstyle{1}$}
\put(-16.6,18.3){$\scriptstyle{1}$}
\put(-11,9){$\scriptstyle{0}$}
\put(3,22){$\scriptstyle{1}$}
\put(5,5.5){$\scriptstyle{0}$}
\put(25,15){$\scriptstyle{0}$}
\put(-80,-23){$\OK \otimes \ok^2 \otimes \OK
\otimes \ap \otimes \Am \otimes \ok$}
}
\put(280,0){
\put(-88,97){$\scriptstyle{0}$}
\put(-121,78){$\scriptstyle{0}$}
\put(-124,59){$\scriptstyle{1}$}
\put(92,117){$\scriptstyle{1}$}
\put(92,105){$\scriptstyle{1}$}
\put(94,10){$\scriptstyle{1}$}
\put(9,42){$\scriptstyle{1}$}
\put(28,33){$\scriptstyle{1}$}
\put(-30,0){\rotatebox[origin=l]{-42}{\line(-1,0){114}}}
\put(-37.5,-1){\rotatebox[origin=l]{42}{\vector(1,0){159}}}
\put(-5.5,0){\rotatebox[origin=l]{-51.06}{\line(-1,0){122}}}
\put(-14,-2){\rotatebox[origin=l]{51.06}{\vector(1,0){149}}}
\put(60,0){\rotatebox[origin=l]{-18.74}{\line(-1,0){187}}}
\put(56.5,0){\rotatebox[origin=l]{18.06}{\vector(1,0){34}}}
\drawline(-120,0)(95,0)
\multiput(60,0)(0,2){40}{\put(0,0){\line(0,1){0.6}}}
\multiput(-5.45455,0)(0,2){11}{\put(0,0){\line(0,1){0.6}}}
\multiput(-30,0)(0,2){15}{\put(0,0){\line(0,1){0.6}}}
\put(-65,45.6){$\scriptstyle{1}$}
\put(-71,25){$\scriptstyle{0}$}
\put(-21,30){$\scriptstyle{1}$}
\put(-27,15){$\scriptstyle{0}$}
\put(-21,2){$\scriptstyle{1}$}
\put(-16.6,18.3){$\scriptstyle{1}$}
\put(-11,9){$\scriptstyle{0}$}
\put(3,22){$\scriptstyle{1}$}
\put(5,5.5){$\scriptstyle{1}$}
\put(25,15){$\scriptstyle{1}$}
\put(-80,-23){$\OK \otimes \ok^2 \otimes \OK 
\otimes \ok \otimes \ichi \otimes \am$}
}
\end{picture}
\end{align*}
\setlength\unitlength{1pt}

The the top left diagram yields the LHS while the other ones lead to the RHS.
We have also shown the corresponding $q$-boson valued amplitude.
As the result the quantized $G_2$ reflection equation (\ref{qre})
in this case becomes the following equation in 
$\mathrm{End}({F}_{q^3} 
\otimes {F}_{q} 
\otimes {F}_{q^3} 
\otimes {F}_{q}
\otimes {F}_{q^3} 
\otimes {F}_{q})$:
\begin{equation}\label{eqd16}
\begin{split}
(\ichi \otimes \ok \otimes \OK 
\otimes \ok^2 \otimes \OK \otimes \am)\mathscr{F}
=&\mathscr{F}(\Ap \otimes \am \otimes \OK 
\otimes \ok^2 \otimes \OK \otimes \ok
+\OK \otimes (\ap)^2 \otimes \Am 
\otimes \ok^2 \otimes \OK \otimes \ok \\
&+r \OK \otimes \ok\,\ap \otimes \ichi 
\otimes \ok\,\am \otimes \OK \otimes \ok
+\OK \otimes \ok^2 \otimes \Ap 
\otimes (\am)^2 \otimes \OK \otimes \ok \\
&+\OK \otimes \ok^2 \otimes \OK
\otimes \ap \otimes \Am \otimes \ok
+\OK \otimes \ok^2 \otimes \OK 
\otimes \ok \otimes \ichi \otimes \am),
\end{split}
\end{equation}
where we have combined 
the terms with coefficients $u_1u_2$ and $u_3u_4$ 
together by means of (\ref{rdef}).

The quantized $G_2$ reflection equation (\ref{qre}) is the set 
of $2^6=64$ equations like (\ref{eqd16})  
corresponding to the choice of $a,b,c,i,j,k \in \{0,1\}$ in (\ref{minami}).
In the next section they will be identified with
the intertwining relation of certain $A_q(G_2)$ modules. 

\section{$A_q(G_2)$ and its intertwiner}\label{sec:rk}

\subsection{\mathversion{bold}Intertwining relation of $A_q(G_2)$ modules}
The quantized coordinate ring $A_q(G_2)$ is a Hopf algebra which can be realized 
by 49 generators $(t_{i,j})_{1\le i,j \le 7}$ obeying 
the so-called $RTT$ type quadratic relations and some additional ones.
They are available in \cite{Sasaki}, which was 
adopted in \cite[Sec.3.3.3]{KOY} in the form directly relevant to this paper. 
Their concrete form is not necessary here.
What we need is the two fundamental representations  
$\pi_i: A_q(G_2) \rightarrow \mathrm{End}(F_{q_i})$ 
associated with the simple reflections $s_i$ $(i=1,2)$,
where $q_1=q, q_2=q^3$.
\begin{equation}\label{pidef}
\begin{split}
&\bigl(\pi_1(t_{i,j})\bigr)_{1 \le i,j \le 7}=
\begin{pmatrix}
\am &  \ok &  0 & 0 & 0 & 0 &0 \\
- \ok & \ap & 0 & 0&0 & 0& 0\\
0&0 & (\am)^2 & 
r\ok\,\am & \ok^2 & 0& 0\\
0 & 0& -\am \, \ok & 
\sk & \ok\,\ap & 0& 0\\
0& 0& \ok^2 & -r \ok\,\ap & (\ap)^2 & 0& 0\\
0&0&0&0&0& \am & \ok \\
0&0&0&0&0& -\ok & \ap
 \end{pmatrix}, \\
&\bigl(\pi_2(t_{i,j})\bigr)_{1 \le i,j \le 7}=
\begin{pmatrix}
1&  0&  0& 0& 0& 0&0\\
 0& \Am & \OK &0 &0 &0 &0 \\
 0& -\OK & \Ap & 0& 0& 0&0\\
0 &0 & 0& 1& 0&0 &0\\
 0& 0& 0& 0& \Am & \OK & 0\\
0 & 0& 0& 0& -\OK & \Ap & 0\\
 0&0&0&0&0&0& 1
 \end{pmatrix}.
\end{split}
\end{equation}
Here $r$ and $\sk$ are defined in (\ref{rdef}) and (\ref{spd}).
These expressions are obtained from \cite[eq.(27)]{KOY}\footnote{
Note that unlike (\ref{yum}) in this paper, 
the operator $\ok$ in \cite[eq.(17)]{KOY} 
does {\em not} contain the zero point energy.}  
by setting $\mu_1= \mu_2=1$,
$\alpha_1=q^{\hf}$, $\alpha_2=q^{\thf}$.

The coproduct 
$\Delta^{(k)} : A_q(G_2) \rightarrow 
A_q(G_2)^{\otimes k}$ takes the simple form
$\Delta^{(k)} (t_{i,j}) = \sum_{1 \le l_2,\ldots, l_k \le 7}
 t_{i, l_2} \otimes t_{l_2,l_3} \otimes \cdots \otimes t_{l_k,j}$. 
For $i_1, \ldots, i_k \in \{1,2\}$,
one can construct a tensor product representation by
\[
\pi_{i_1,\ldots, i_k}:= 
(\pi_{i_1} \otimes \cdots \otimes \pi_{i_k}) \circ \Delta^{(k)}:\, 
A_q(G_2) \rightarrow \mathrm{End}(F_{q_{i_1}}\otimes \cdots \otimes F_{q_{i_k}}).
\]
According to the general theory \cite{So2},
$\pi_{i_1,\ldots, i_k}$ is irreducible if and only if 
$s_{i_1}\cdots s_{i_k}$ is a reduced expression of an element of 
the Weyl group of $G_2$.
Moreover $\pi_{i_1,\ldots, i_k}$ and $\pi_{j_1,\ldots, j_k}$
are equivalent if $s_{i_1}\cdots s_{i_k} = s_{j_1}\cdots s_{j_k}$.
We are concerned with the two reduced expressions of the longest element 
$s_2s_1s_2s_1s_2s_1 = s_1s_2s_1s_2s_1s_2$, 
the associated representations $\pi_{212121}$ and $\pi_{121212}$
and their isomorphism $\pi_{212121} \simeq \pi_{121212}$.

Let $\Phi^\vee$ be the intertwiner. 
Namely it is the map 
$F_{q}\otimes F_{q^3} \otimes F_{q}\otimes F_{q^3} \otimes
F_{q}\otimes F_{q^3} \rightarrow 
F_{q^3} \otimes F_{q}\otimes F_{q^3} \otimes
F_{q}\otimes F_{q^3} \otimes F_{q}$
characterized by $\pi_{212121} \Phi^\vee = \Phi^\vee \pi_{121212}$ up to 
normalization.
Set $\Phi = \Phi^\vee \circ P$ where $P$ is a linear map 
reversing the order of the six-fold tensor product as 
$P(x_1\otimes x_2 \otimes \cdots \otimes x_6) = 
x_6 \otimes x_5 \cdots \otimes x_1$.  
Thus there exists the unique $\Phi$ such that 
\begin{align}
&\Phi \in \mathrm{End}(
F_{q^3} \otimes F_{q}\otimes F_{q^3} \otimes
F_{q}\otimes F_{q^3} \otimes F_{q}),\\
&\pi_{212121}(g) \,\Phi  = \Phi \, \pi'_{121212}(g)\quad 
\forall g \in A_q(G_2) \quad\qquad 
(\pi'_{121212} := P \pi_{121212} P),
\label{FS}\\
&\Phi (|0 \rangle \otimes |0 \rangle \otimes
|0 \rangle \otimes|0 \rangle\otimes|0 \rangle\otimes|0 \rangle) = 
|0 \rangle \otimes |0 \rangle \otimes
|0 \rangle \otimes|0 \rangle
\otimes|0 \rangle\otimes|0 \rangle.
\label{nor}
\end{align}
The condition (\ref{nor}) fixes the normalization.
It suffices to impose the equation (\ref{FS}) 
for the 49 generators $g=t_{i,j}$. 
By using the explicit form of the coproduct $\Delta^{(6)}$, 
they are expressed as
\begin{equation}\label{kkna}
\begin{split}
&\left(\sum \pi_2(t_{i,l_2}) 
\otimes \pi_1(t_{l_2, l_3})
\otimes \pi_2(t_{l_3, l_4})
\otimes \pi_1(t_{l_4, l_5})
\otimes\pi_2(t_{l_5, l_6})
\otimes \pi_1(t_{l_6, j})\right)\Phi  \\
&= \Phi \left(\sum
\pi_2(t_{l_6,j})
\otimes \pi_1(t_{l_5, l_6})
\otimes \pi_2(t_{l_4, l_5})
\otimes \pi_1(t_{l_3, l_4})
\otimes \pi_2(t_{l_2, l_3})
\otimes \pi_1(t_{i, l_2})
\right)\quad (1 \le i, j \le 7),
\end{split}
\end{equation}
where the sums are taken over $1 \le l_2,\ldots, l_6 \le 7$.
In this way the intertwining relation (\ref{FS}) 
boils down to the 49 equations (\ref{kkna}).
Although the lists of $\pi_1(t_{i,j}), \pi_2(t_{i,j})$ in (\ref{pidef})
are pretty sparse, some equations become lengthy including 
typically 16 terms on one side or both.
We do not display them all here but present a few examples. 
\begin{align}
g = &\, t_{1,1}: \nonumber\\
&(\ichi\otimes \am\otimes \ichi\otimes \am
\otimes \ichi\otimes \am - 
 \ichi\otimes \am\otimes \ichi\otimes \ok
\otimes \Am\otimes \ok - 
 \ichi\otimes \ok\otimes \Am\otimes \ap
\otimes \Am\otimes \ok  \nonumber\\
&\qquad-\ichi\otimes \ok\otimes \Am
\otimes \ok\otimes \ichi\otimes \am + 
 \ichi\otimes \ok\otimes \OK
\otimes (\am)^2 \otimes \OK\otimes \ok) \Phi
\nonumber\\
&= \Phi(\ichi\otimes \am\otimes \ichi
\otimes \am\otimes \ichi\otimes \am - 
 \ichi\otimes \am\otimes \ichi\otimes \ok\otimes \Am\otimes \ok - 
 \ichi\otimes \ok\otimes \Am
\otimes \ap\otimes \Am\otimes \ok \nonumber\\
&\qquad -\ichi\otimes \ok\otimes \Am
\otimes \ok\otimes \ichi\otimes \am + 
 \ichi\otimes \ok\otimes \OK\otimes 
(\am)^2 \otimes \OK\otimes \ok),\nonumber\\
g = & \, t_{1,5}:\nonumber\\
&(\ichi \otimes \am \otimes \ichi 
\otimes \ok \otimes \OK \otimes \ok^2 + 
  \ichi \otimes \ok \otimes \Am 
\otimes \ap \otimes \OK \otimes \ok^2 + 
  \ichi \otimes \ok \otimes \OK 
\otimes (\am)^2 \otimes \Ap \otimes \ok^2 \nonumber\\
& \quad + 
r \ichi \otimes \ok \otimes \OK 
\otimes \ok\, \am \otimes \ichi \otimes \ok\, \ap + 
  \ichi \otimes \ok \otimes \OK 
\otimes \ok^2 \otimes \Am \otimes (\ap)^2)\Phi
\nonumber\\
& =\Phi(\Am \otimes (\ap)^2 \otimes \Am 
\otimes \ok^2 \otimes \OK \otimes \ok + 
  r \Am \otimes \ok\, \ap \otimes \ichi 
\otimes \ok\, \am \otimes \OK \otimes \ok + 
  \Am \otimes \ok^2 \otimes \Ap 
\otimes (\am)^2 \otimes \OK \otimes \ok \nonumber\\
& \quad+ 
  \Am \otimes \ok^2 \otimes \OK 
\otimes \ap \otimes \Am \otimes \ok + 
  \Am \otimes \ok^2 \otimes \OK 
\otimes \ok \otimes \ichi \otimes \am - 
  \OK \otimes \am \otimes \OK 
\otimes \ok^2 \otimes \OK \otimes \ok),\nonumber\\
g = & \, t_{1,6}:\nonumber\\
&(\ichi\otimes \ok\otimes \OK\otimes \ok^2 \otimes \OK\otimes \am)\Phi
\nonumber\\
&= \Phi(
\Ap\otimes \am\otimes \OK\otimes \ok^2\otimes \OK\otimes \ok + 
 \OK\otimes (\ap)^2\otimes \Am
\otimes \ok^2 \otimes \OK\otimes \ok +  
 r \OK\otimes \ok\, \ap\otimes \ichi
\otimes \ok \, \am\otimes \OK\otimes \ok
\nonumber\\
& \qquad + \OK\otimes \ok^2\otimes \Ap
\otimes (\am)^2\otimes \OK\otimes \ok + 
 \OK\otimes \ok^2\otimes \OK\otimes \ap
\otimes \Am\otimes \ok + 
 \OK\otimes \ok^2\otimes \OK\otimes \ok\otimes \ichi\otimes \am),
 \label{eq16}\\
g = & \, t_{2,6}:\nonumber\\
&({\Am}\otimes{\ap}\otimes\OK\otimes\ok^2\otimes\OK\otimes{\am} + 
 \OK\otimes(\am)^2 \otimes{\Ap}\otimes\ok^2\otimes\OK\otimes{\am} + 
 r \OK\otimes{\ok\, \am}\otimes\ichi 
\otimes{\ok\, \ap}\otimes\OK\otimes{\am} 
\nonumber\\
& \qquad+ 
 \OK\otimes\ok^2\otimes{\Am}\otimes(\ap)^2\otimes\OK\otimes{\am} + 
 \OK\otimes\ok^2\otimes\OK\otimes{\am}\otimes{\Ap}\otimes{\am} - 
 \OK\otimes\ok^2\otimes\OK\otimes{\ok}\otimes\ichi \otimes{\ok})\Phi
\nonumber\\
&=  \Phi({\Ap}\otimes{\am}\otimes\OK\otimes\ok^2\otimes\OK\otimes{\ap} + 
 \OK\otimes(\ap)^2\otimes{\Am}\otimes\ok^2\otimes\OK\otimes{\ap} + 
 r \OK\otimes{\ok\, \ap}\otimes\ichi 
\otimes{\ok\, \am}\otimes\OK\otimes{\ap} 
\nonumber\\
& \qquad+ 
 \OK\otimes\ok^2\otimes{\Ap}\otimes(\am)^2 \otimes\OK\otimes{\ap} + 
 \OK\otimes\ok^2\otimes\OK\otimes{\ap}\otimes{\Am}\otimes{\ap} - 
 \OK\otimes\ok^2\otimes\OK\otimes{\ok}\otimes\ichi \otimes{\ok}),
\nonumber\\
g = & \, t_{1,7}, t_{7,1}:\nonumber\\
&[\Phi, \ichi\otimes \ok\otimes \OK\otimes \ok^2 \otimes \OK\otimes \ok]=0.
\nonumber
\end{align}

\subsection{\mathversion{bold}Solution to the quantized $G_2$ reflection equation}

Notice that (\ref{eq16}) coincides with (\ref{eqd16})
under the identification $\mathscr{F} = \Phi$.
In fact under this correspondence 
one can directly check that the 49 intertwining relations (\ref{kkna})
and the 64 quantized $G_2$ reflection equations (\ref{minami}) 
are {\em equivalent}. 
We list the correspondence of the 
indices $(i,j)$ in (\ref{kkna}) and 
$(a,b,c,i,j,k)$ in (\ref{minami}) in Appendix \ref{app:ind}.
Since (\ref{nor}) and (\ref{tsgmi}) impose the same normalization
on $\Phi$ and $\mathscr{F}$, we conclude $\mathscr{F}=\Phi$.
Let us summarize this result in

\begin{theorem}\label{th:sol}
Under the normalization (\ref{tsgmi}),
the quantized $G_2$ reflection equation (\ref{qre}) 
with $L, J$ given in Section \ref{ss:L} and \ref{ss:J}
has the unique solution $\mathscr{F}=\Phi$ 
in terms of the intertwiner $\Phi$ of the $A_q(G_2)$ module
characterized by (\ref{FS}) and (\ref{nor}).
\end{theorem}

Henceforth we shall identify $\Fm$ and $\Phi$ and 
write $\mathscr{F}$ to also mean the intertwiner $\Phi$.
Let us quote some basic properties of $\mathscr{F}$ from 
\cite[Sec.4.4]{KOY}.
Set
\begin{equation*}
\Fm\,( |i\rangle \otimes |j\rangle \otimes |k\rangle \otimes 
|l\rangle \otimes |m\rangle \otimes |n \rangle) =
\sum_{a,b,c,d,e,f \in \Z_{\ge 0}}
\Fm^{abcdef}_{ijklmn}|a\rangle \otimes |b\rangle \otimes
 |c\rangle \otimes |d\rangle \otimes |e\rangle \otimes |f \rangle.
\end{equation*}
Then the following properties are valid:
\begin{align}
&\mathscr{F}^{abcdef}_{ijklmn} \in \mathbb{Z}[q],\;\;
\mathscr{F}^{abcdef}_{ijklmn}=0 \;\;\mathrm{unless}\; \;
\left({a\!+\!b\!+\!2c\!+\!d\!+\!e \atop b\!+\!3c\!+\!2d\!+\!3e\!+\!f}\right)=
\left({i\!+\!j\!+\!2k\!+\!l\!+\!m\atop j\!+\!3k\!+\!2l\!+\!3m\!+\!n}\right),
\label{claw}\\
&\mathscr{F}^{-1} = \mathscr{F},\quad 
\mathscr{F}^{abcdef}_{ijklmn} 
=\frac{(q^6)_i(q^2)_j(q^6)_k(q^2)_l(q^6)_m(q^2)_n}
{(q^6)_a(q^2)_b(q^6)_c(q^2)_d(q^6)_e(q^2)_f}
\mathscr{F}^{ijklmn}_{abcdef}.
\label{fpro}
\end{align}
Due to the latter property of (\ref{claw}), 
$\mathscr{F}$ is an infinite direct sum 
of finite dimensional matrices.
In terms of ${\bf h}_i$ acting as ${\bf h}$ (\ref{syri})
on the $i$ th component from the left,
it may be rephrased as the commutativity 
\begin{equation}\label{noi}
[\Fm, x^{{\bf h}_1} 
(xy)^{{\bf h}_2}
(x^2y^3)^{{\bf h}_3}
(xy^2)^{{\bf h}_4}
(xy^3)^{{\bf h}_5}
y^{{\bf h}_6}]=0,
\end{equation}
where $x$ and $y$ are free parameters.
We let $\Fm$ also act on 
$\langle \omega| \in 
F^\ast_{q^3} \otimes F^\ast_{q}\otimes F^\ast_{q^3} \otimes
F^\ast_{q}\otimes F^\ast_{q^3} \otimes F^\ast_{q}$
by $(\langle \omega | \Fm) |\omega'\rangle = 
\langle \omega | (\Fm |\omega'\rangle)$
for any $| \omega'\rangle \in 
F_{q^3} \otimes F_{q}\otimes F_{q^3} \otimes
F_{q}\otimes F_{q^3} \otimes F_{q}$.

It is possible to make a tedious computer program to 
calculate $\mathscr{F}^{abcdef}_{ijklmn}$ for any given indices 
by using (\ref{kkna}).
However unlike the $A_q(A_2)$ and $A_q(C_2)$ cases,   
an explicit general formula is yet to be constructed.
At $q=0$ $\mathscr{F}^{abcdef}_{ijklmn}$ is known to become $0$ or $1$,
which can be determined 
by the ultradiscretization (tropical form) of \cite[Th.3.1(c)]{BZ97}.

\begin{example}
The following is the list of all the nonzero $\mathscr{F}^{abcdef}_{100102}$.
\begin{alignat*}{2}
\mathscr{F}^{000200}_{100102}
& = -q^3 (1 - q^4)(1-q^6) (1 - q^2 - q^6),
&
\mathscr{F}^{001001}_{100102}
& = -q^2 (1 - q^4)^2(1-q^6) (1 + q^4), \\
\mathscr{F}^{010010}_{100102}
& = (1 - q^4)(1-q^6) (1 - q^2 + q^{10}), 
&
\mathscr{F}^{200004}_{100102}
& = q^{11}, \\
\mathscr{F}^{020002}_{100102}
& =  q^3(1-q^6)(1 - q^4 - q^6 - 2 q^8 - q^{10} - q^{12}), 
&
\mathscr{F}^{100011}_{100102}
& = q^2 (1 - q^4) (1 - q^6 + q^{14}), \\
\mathscr{F}^{100102}_{100102}
& =  q^4 (1 + q^2 - 2 q^6 - q^8 - q^{10} + q^{14} + q^{16} + q^{18}), 
&\mathscr{F}^{110003}_{100102}
& = q^6(1+q^2)(1-q^8-q^{12}), \\
\mathscr{F}^{010101}_{100102}
& =  q(1+q^2)(1-q^6)(1 - q^2 - q^4 - q^6 + q^{10} + q^{12} + q^{14}).
\end{alignat*}
\end{example}

\section{Matrix product solutions to Yang-Baxter equation from $L$}\label{sec:ybe}

\subsection{Tetrahedron equation}

The $q$-boson valued $L$ matrix (\ref{Lop}) is known to satisfy 
a version the tetrahedron equation \cite{BS}
\begin{align}
&L_{124}L_{135} L_{236} \,\Rm_{456} = 
\Rm_{456}L_{236}L_{135}L_{124} 
\in \mathrm{End}(
\overset{1}{V}\otimes 
\overset{2}{V}\otimes 
\overset{3}{V}  
\otimes \overset{4}{F}_{q^3} \otimes \overset{5}{F}_{q^3} 
\otimes \overset{6}{F}_{q^3}).
\label{LLLR}
\end{align}
This is a Yang-Baxter equation up to conjugation by 
$\Rm \in \mathrm{End}(F_{q^3}\otimes F_{q^3}\otimes F_{q^3})$.
Such an $\Rm$ is unique up to overall normalization 
and is known to satisfy 
the tetrahedron equation 
$\Rm_{124}\Rm_{135} \Rm_{236} \,\Rm_{456} = 
\Rm_{456}\Rm_{236}\Rm_{135}\Rm_{124}$ among themselves.
See \cite{KV} for the approach 
from the representation theory of the quantized coordinate ring $A_q(A_2)$,  
\cite{BS} for a quantum geometry argument and \cite[Sec.3.1]{KP}
for a brief guide to the background.

We let $\Rm$ also act on $(F_{q^3}^\ast)^{\otimes 3}$ by 
$\bigl((\langle a| \otimes \langle b| \otimes \langle c| )\Rm\bigr) 
(|i\rangle \otimes |j\rangle \otimes |k\rangle)
=   (\langle a| \otimes \langle b| \otimes \langle c| )
\bigl(\Rm( |i\rangle \otimes |j\rangle \otimes |k\rangle)\bigr)$.
In this paper we will only need the following properties 
\begin{align}
&\Rm^{-1}=\Rm,\qquad
[\Rm_{123}, x^{{\bf h}_1+{\bf h}_2}y^{{\bf h}_2+{\bf h}_3}]=0,
\label{ykn3}
\\
&(\langle \chi| \otimes \langle \chi |\otimes \langle\chi|) \Rm = 
\langle \chi| \otimes \langle \chi |\otimes \langle \chi|,\quad
\Rm (|\chi\rangle \otimes|\chi\rangle \otimes|\chi\rangle)
=  |\chi\rangle \otimes|\chi\rangle \otimes|\chi\rangle,
\label{tbsa}
\\
&\langle \chi | = \sum_{m \ge 0} \frac{\langle m|}{(q^3)_m} \in 
F^\ast_{q^3},\qquad
|\chi\rangle = \sum_{m \ge 0} \frac{|m\rangle}{(q^3)_m}
\in F_{q^3},
\label{bve}
\end{align}
where $x,y$ are free parameters.
The relation (\ref{tbsa}) was proved in \cite[Pr.4.1]{KS}\footnote{
$\chi$ is just a symbol 
and not a nonnegative integer. A similar caution applies also to
$\xi$ in (\ref{bve2}).}.

\subsection{Concatenation of the tetrahedron equation}
Consider $n$ copies of (\ref{LLLR}) in which the 
spaces labeled with $1,2,3$ are replaced by $1_i,2_i,3_i$ with 
$i=1,2,\ldots, n$:
\begin{align*}
&(L_{1_i2_i4}L_{1_i3_i5} 
L_{2_i3_i6}) \,\Rm_{456} = 
\Rm_{456}\,(L_{2_i3_i6}L_{1_i3_i5}L_{1_i2_i4}).
\end{align*}
Sending $\Rm_{456}$ to the left by 
repeatedly applying this relation, we get
\begin{equation}\label{mikrup}
\begin{split}
&(L_{1_1 2_1 4}L_{1_1 3_1 5} 
L_{2_1 3_1 6})\cdots 
(L_{1_n 2_n 4}L_{1_n 3_n 5} 
L_{2_n 3_n 6})\,\Rm_{456} \\
&\qquad = \Rm_{456} \,(L_{2_13_16}
L_{1_13_15}
L_{1_12_14}) \cdots 
(L_{2_n 3_n 6}L_{1_n 3_n 5}
L_{1_n 2_n 4}).
\end{split}
\end{equation}
Set ${\bf V} = V^{\otimes n} \simeq (\C^2)^{\otimes n}$ in general and  
$\overset{k}{\bf V} 
= \overset{k_1}{V} \otimes \cdots \otimes \overset{k_n}{V}$
when the label is present.
The equality (\ref{mikrup}) holds in  
$\mathrm{End}(\overset{1}{\bf V} \otimes
\overset{2}{\bf V} \otimes
\overset{3}{\bf V} \otimes \overset{4}{F}_{q^3}
\otimes \overset{5}{F}_{q^3}
\otimes \overset{6}{F}_{q^3})$.
It is possible to rearrange it 
without changing the order of any two operators 
sharing common labels as
\begin{equation}\label{mikru0}
\begin{split}
&(L_{1_1 2_1 4} \cdots L_{1_n 2_n 4})
(L_{1_1 3_1 5}  \cdots L_{1_n 3_n 5})
(L_{2_1 3_1 6}  \cdots L_{2_n 3_n 6})
\,\Rm_{456} \\
&\qquad = \Rm_{456} \,
(L_{2_1 3_1 6}  \cdots L_{2_n 3_n 6})
(L_{1_1 3_1 5}  \cdots L_{1_n 3_n 5})
(L_{1_1 2_1 4} \cdots L_{1_n 2_n 4}).
\end{split}
\end{equation}
Write the right relation in (\ref{ykn3}) as  
$\Rm_{456}^{-1}x^{{\bf h}_4}(xy)^{{\bf h}_5} y^{{\bf h}_6} 
= x^{{\bf h}_4}(xy)^{{\bf h}_5} y^{{\bf h}_6} \Rm_{456}^{-1}$.
Multiplying this to (\ref{mikru0}) from the left we get 
\begin{equation}\label{mikru}
\begin{split}
&\Rm^{-1}_{456}
\bigl(x^{{\bf h}_4}L_{1_1 2_1 4} \cdots L_{1_n 2_n 4}\bigr)
\bigl((xy)^{{\bf h}_5}L_{1_1 3_1 5}  \cdots L_{1_n 3_n 5}\bigr)
\bigl(y^{{\bf h}_6}L_{2_1 3_1 6}  \cdots L_{2_n 3_n 6}\bigr)
\,\Rm_{456} \\
&\qquad = 
\bigl(y^{{\bf h}_6}L_{2_1 3_1 6}  \cdots L_{2_n 3_n 6}\bigr)
\bigl((xy)^{{\bf h}_5}L_{1_1 3_1 5}  \cdots L_{1_n 3_n 5}\bigr)
\bigl(x^{{\bf h}_4}L_{1_1 2_1 4} \cdots L_{1_n 2_n 4}\bigr).
\end{split}
\end{equation}

\subsection{Reduction to Yang-Baxter equation}
The trace of (\ref{mikru}) over 
$\overset{4}{F}_{q^3}
\otimes \overset{5}{F}_{q^3}
\otimes \overset{6}{F}_{q^3}$ gives
\begin{equation}\label{mikru1}
\begin{split}
&\mathrm{Tr}_4(x^{{\bf h}_4} L_{1_1 2_1 4}\cdots
L_{1_n 2_n 4})
\mathrm{Tr}_5((xy)^{{\bf h}_5} L_{1_1 3_1 5}\cdots
L_{1_n 3_n 5})
\mathrm{Tr}_6(y^{{\bf h}_6} L_{2_1 3_1 6}\cdots
L_{2_n 3_n 6})\\
&= \mathrm{Tr}_6(y^{{\bf h}_6} L_{2_1 3_1 6}\cdots
L_{2_n 3_n 6})
\mathrm{Tr}_5((xy)^{{\bf h}_5} L_{1_1 3_1 5}\cdots
L_{1_n 3_n 5})
\mathrm{Tr}_4(x^{{\bf h}_4} L_{1_1 2_1 4}\cdots
L_{1_n 2_n 4}).
\end{split}
\end{equation}
Alternatively one may sandwich (\ref{mikru}) 
between the bra vector 
$(\langle \overset{4}{\chi}| \otimes
\langle \overset{5}{\chi}| \otimes
\langle \overset{6}{\chi}|)$ 
and the ket vector 
$|\overset{4}{\chi}\rangle 
\otimes |\overset{5}{\chi}\rangle  \otimes 
|\overset{6}{\chi}\rangle$.
From $\Rm^{-1}=\Rm$ (\ref{ykn3}) and (\ref{tbsa}), the result becomes
\begin{equation}\label{mikru2}
\begin{split}
&\langle\overset{4}{\chi}| x^{{\bf h}_4}
L_{1_1 2_1 4}\cdots
L_{1_n 2_n 4}|\overset{4}{\chi}\rangle
\langle\overset{5}{\chi}| (xy)^{{\bf h}_5}
L_{1_1 3_1 5}\cdots
L_{1_n 3_n 5}|\overset{5}{\chi}\rangle
\langle\overset{6}{\chi}| y^{{\bf h}_6}
L_{2_1 3_1 6}\cdots
L_{2_n 3_n 6}|\overset{6}{\chi}\rangle\\
&=
\langle\overset{6}{\chi}| y^{{\bf h}_6}
L_{2_1 3_1 6}\cdots
L_{2_n 3_n 6}|\overset{6}{\chi}\rangle
\langle\overset{5}{\chi}| (xy)^{{\bf h}_5}
L_{1_1 3_1 5}\cdots
L_{1_n 3_n 5}|\overset{5}{\chi}\rangle
\langle\overset{4}{\chi}| x^{{\bf h}_4}
L_{1_1 2_1 4}\cdots
L_{1_n 2_n 4}|\overset{4}{\chi}\rangle.
\end{split}
\end{equation}

Set
\begin{align}\label{obata1}
R^{\mathrm {tr}}_{1,2}(z) 
&= \varrho^{\mathrm {tr}}(z)
\mathrm{Tr}_a(z^{{\bf h}_a} L_{1_1 2_1 a}\cdots
L_{1_n 2_n a}) \in 
\mathrm{End}(\overset{1}{\bf V} \otimes \overset{2}{\bf V}),
\\
R^{\mathrm{bv}}_{1, 2}(z) 
&= \varrho^{\mathrm{bv}}(z)
\langle \overset{a}{\chi}| 
z^{{\bf h}_a} L_{1_1 2_1 a}\cdots
L_{1_n 2_n a}
|\overset{a}{\chi} \rangle\in 
\mathrm{End}(\overset{1}{\bf V} \otimes \overset{2}{\bf V}),
\label{obata2}
\end{align}
where $a$ is a dummy label for the auxiliary Fock space $\overset{a}{F}_{q^3}$.
The normalization factors $\varrho^{\mathrm {tr}}(z)$ and 
$\varrho^{\mathrm {br}}(z)$ will be specified in (\ref{askS}).
Now (\ref{mikru1}) and (\ref{mikru2}) are both stated as the Yang-Baxter equation
\begin{equation}\label{ybe1}
R_{12}(x)
R_{13}(xy)
R_{23}(y)
=
R_{23}(y)
R_{13}(xy)
R_{12}(x)
\end{equation}
with $R(z) = R^{\mathrm {tr}}(z)$ and $R^{\mathrm {bv}}(z)$.
We call the above procedure to get the solutions
$R^{\mathrm {tr}}(z)$ and $R^{\mathrm {bv}}(z)$ of the 
Yang-Baxter equation from the tetrahedron equation (\ref{LLLR})
the {\em trace reduction} and the {\em boundary vector reduction}, respectively.
The vectors (\ref{bve}) are referred to as {\em boundary vectors}.

The trace reduction is due to \cite{BS}
and the boundary vector reduction in this paper 
is a special case of more general ones in \cite{KS}.
The solutions  $R^{\mathrm {tr}}(z)$ and $R^{\mathrm {bv}}(z)$
have been identified with the quantum $R$ matrices 
for the antisymmetric tensor representations of $U_p(A^{(1)}_{n-1})$
and the spin representation of $U_p(D^{(2)}_{n+1})$ with $p^2=-q^{-3}$.
A concise summary of these results can be found in 
\cite[App.B]{KP}\footnote{$R^{\mathrm {bv}}(z)$
corresponds to $S^{1,1}(z)$ in \cite{KP}.
Since $L$ in \cite{KP} is $q^2$-boson valued, 
the formulas in \cite[App.B]{KP} fit this paper
if $q$ there is replaced by $q^\thf$ .}.

\subsection{\mathversion{bold}Basic properties of 
$R^{\mathrm {tr}}(z)$ and $R^{\mathrm {bv}}(z)$}\label{ss:bss}

We write the base vectors of ${\bf V} = V^{\otimes n}$ as
$|\alb\rangle= v_{\alpha_1} \otimes \cdots \otimes v_{\alpha_n}$
in terms of an array 
$\alb=(\alpha_1,\ldots,\alpha_n) \in \{0,1\}^n$.
We warn that $|\alb\rangle \in {\bf V}$ should not be confused with 
the base $|m\rangle$ of a Fock space
containing a single nonnegative integer. 
Set
\begin{align*}
&R(z)(|\alb\rangle \otimes |\beb\rangle)
= \sum_{\gab,\deb \in \{0,1\}^n}
R(z)_{\alb,\beb}^{\gab,\deb}
|\gab\rangle \otimes |\deb\rangle\qquad 
(R = R^{\mathrm {tr}}, R^{\mathrm {bv}}).
\end{align*}
Then (\ref{obata1}) and (\ref{obata2}) 
imply the matrix product formulas as
\begin{align}
&R^{\mathrm {tr}}(z)_{\alb,\beb}^{\gab,\deb}
= \varrho^{\mathrm {tr}}(z)
\mathrm{Tr}\bigl(z^{{\bf h}}
L^{\gamma_1,\delta_1}_{\alpha_1, \beta_1}
\cdots
L^{\gamma_n,\delta_n}_{\alpha_n, \beta_n}\bigr),
\label{strz}
\\
&R^{\mathrm {bv}}(z)_{\alb,\beb}^{\gab,\deb}
= \varrho^{\mathrm {bv}}(z)
\langle \chi |z^{{\bf h}}
L^{\gamma_1,\delta_1}_{\alpha_1, \beta_1}
\cdots
L^{\gamma_n,\delta_n}_{\alpha_n, \beta_n}|\chi\rangle,
\label{s11}
\end{align}
where $L^{\gamma,\delta}_{\alpha,\beta}$ is 
given by (\ref{Lop}) and $\mathrm{Tr}(\cdots)$ and 
$\langle \chi| (\cdots) |\chi\rangle$ are taken over $F_{q^3}$.
They are evaluated by using the commutation relations (\ref{ngh2}), 
the formula $(\ref{lin})|_{q \rightarrow q^3}$ and 
\begin{align}
&\mathrm{Tr}(z^{\bf h} \OK^r(\Ap)^s(\Am)^{s'}) = 
\delta_{s,s'}\frac{q^{\frac{3r}{2}}(q^6;q^6)_s}{(zq^{3r};q^6)_{s+1}}.
\label{yuk1}
\end{align}

For $\alb=(\alpha_1,\ldots, \alpha_n) \in \{0,1\}^n$ set 
\begin{align}\label{askB}
|\alb| =\alpha_1+\cdots + \alpha_n,\qquad
{\bf V}_k = \bigoplus_{\alb \in \{0,1\}^n, \, |\alb | =k}\!\!\!
\C |\alb\rangle.
\end{align}
By the definition the direct sum decomposition
$
{\bf V}= {\bf V}_0 \oplus {\bf V}_1 \oplus \cdots \oplus {\bf V}_n
$
holds.
From (\ref{mzsma1}), (\ref{mzsma2}) 
and (\ref{bve}) one can show
\begin{align}
&R(z)_{\alb,\beb}^{\gab,\deb}=0 \;\; \text{unless}\;\;
\alb + \beb = \gab + \deb \in \Z^n
\qquad (R= R^{\mathrm {tr}}, R^{\mathrm{bv}}),
\label{yume2}\\
&R^{\mathrm {tr}}(z)_{\alb,\beb}^{\gab,\deb}=0 
\;\; \text{unless}
\;\; |\alb| = |\gab| \;\; 
\text{and}\;\; |\beb| = |\deb|.
\label{yume3}
\end{align}
The property (\ref{yume3}) implies the decomposition
\begin{align}
R^{\mathrm {tr}}(z) &= \bigoplus_{0 \le l,m \le n}
R^{\mathrm {tr}}_{l,m}(z),\qquad\;\,
R^{\mathrm {tr}}_{l,m}(z) \in \mathrm{End}({\bf V}_l\otimes {\bf V}_m).
\label{nami}
\end{align}
The Yang-Baxter equation (\ref{ybe1}) 
with $R(z)=R^{\mathrm {tr}}(z)$ is valid 
for each subspace 
${\bf V}_k \otimes {\bf V}_l \otimes {\bf V}_m$ of 
$\overset{1}{\bf V} \otimes 
\overset{2}{\bf V} \otimes
\overset{3}{\bf V}$.
The scalar prefactor in (\ref{obata1}) for the summand 
$R^{\mathrm {tr}}_{l,m}(z)$ in (\ref{nami}) may be taken as
$\varrho^{\mathrm {tr}}_{l,m}(z)$ depending on $l$ and $m$.
We choose it and the one in (\ref{obata2}) as
\begin{equation}\label{askS}
\varrho^{\mathrm {tr}}_{l,m}(z) =
q^{-\thf |l-m|}(1-z q^{3|l-m|}),\qquad
\varrho^{\mathrm{bv}}(z) = \frac{(z;q^3)_\infty}{(-z q^3; q^3)_\infty}.
\end{equation}
They make all the matrix elements of 
$R^{\mathrm {tr}}_{l,m}(z)$ and $R^{\mathrm{bv}}(z)$ 
rational in $z$ and $q^\hf$.
For example we have 
\begin{equation}\label{askS1}
\begin{split}
&R^{\mathrm {tr}}_{l,m}(z) (|{\bf e}_{[1,l]}\rangle
\otimes |{\bf e}_{[1,m]}\rangle)
=(-1)^{\max(l-m,0)}|{\bf e}_{[1,l]}\rangle
\otimes |{\bf e}_{[1,m]}\rangle,
\\
&R^{\mathrm{bv}}(z) (|{\bf 0} \rangle \otimes |{\bf 0} \rangle)
= |{\bf 0} \rangle \otimes |{\bf 0} \rangle,
\end{split}
\end{equation}
where 
${\bf e}_{[1,k]}
={\bf e}_1+\cdots + {\bf e}_k$ and 
$|{\bf 0} \rangle=|0,0,\ldots, 0\rangle$.

\section{Reduction of quantized $G_2$ reflection equation}\label{sec:re}

Starting from the quantized $G_2$ reflection equation (\ref{hrk}),
one can perform two kinds of reductions similar to Section \ref{sec:ybe} to 
construct solutions to the $G_2$ reflection equation (\ref{reika}) 
in the matrix product form.
This is the main result of the paper which we are going to present in this section.

\subsection{\mathversion{bold}Concatenation of quantized $G_2$ reflection equation}
\label{ss:cc}

Consider $n$ copies of (\ref{hrk}) in which the spaces labeled with $1,2,3$ are
replaced by $1_i, 2_i,3_i$ with $i=1,2,\ldots, n$:
\begin{equation}\label{kanon}
(L_{1_i 2_i 4}J_{1_i 3_i 2_i 5}L_{2_i 3_i 6}
J_{2_i 1_i 3_i 7}L_{3_i 1_i 8}J_{3_i 2_i 1_i 9})
\Fm_{456789}
=
\Fm_{456789}
(J_{2_i 3_i 1_i 9}L_{1_i 3_i 8}J_{1_i 2_i 3_i 7}
L_{3_i 2_i 6}J_{3_i 1_i 2_i 5}L_{2_i 1_i 4}).
\end{equation}
Using (\ref{kanon}) successively, one can 
bring $\Fm_{456789}$ to the left to derive 
\begin{equation*}
\begin{split}
&(L_{1_1 2_1 4}J_{1_1 3_1 2_1 5}L_{2_1 3_1 6}
J_{2_1 1_1 3_1 7}L_{3_1 1_1 8}J_{3_1 2_1 1_1 9})\cdots 
(L_{1_n 2_n 4}J_{1_n 3_n 2_n 5}L_{2_n 3_n 6}
J_{2_n 1_n 3_n 7}L_{3_n 1_n 8}J_{3_n 2_n 1_n 9}) 
\Fm_{456789}\\
&=
\Fm_{456789}
(J_{2_1 3_1 1_1 9}L_{1_1 3_1 8}J_{1_1 2_1 3_1 7}
L_{3_1 2_1 6}J_{3_1 1_1 2_1 5}L_{2_1 1_1 4}) \cdots
(J_{2_n 3_n 1_n 9}L_{1_n 3_n 8}J_{1_n 2_n 3_n 7}
L_{3_n 2_n 6}J_{3_n 1_n 2_n 5}L_{2_n 1_n 4}).
\end{split}
\end{equation*}
This can be rearranged without changing the order of operators 
sharing common labels as
\begin{equation}\label{kanon1}
\begin{split}
&(L_{1_1 2_1 4} \cdots L_{1_n 2_n 4})
(J_{1_1 3_1 2_1 5} \cdots J_{1_n 3_n 2_n 5})
(L_{2_1 3_1 6} \cdots L_{2_n 3_n 6})\\
&\quad\times 
(J_{2_1 1_1 3_1 7} \cdots J_{2_n 1_n 3_n 7})
(L_{3_1 1_1 8} \cdots L_{3_n 1_n 8})
(J_{3_1 2_1 1_1 9} \cdots J_{3_n 2_n 1_n 9}) \Fm_{456789}\\
&=
\Fm_{456789}
(J_{2_1 3_1 1_1 9} \cdots J_{2_n 3_n 1_n 9})
(L_{1_1 3_1 8} \cdots L_{1_n 3_n 8})
(J_{1_1 2_1 3_1 7} \cdots J_{1_n 2_n 3_n 7})\\
&\quad\times
(L_{3_1 2_1 6} \cdots L_{3_n 2_n 6})
(J_{3_1 1_1 2_1 5} \cdots J_{3_n 1_n 2_n 5})
(L_{2_1 1_1 4} \cdots L_{2_n 1_n 4}).
\end{split}
\end{equation}
Write (\ref{noi}) as 
\begin{align*}
\Fm^{-1}_{456789}
x^{{\bf h}_4}(xy)^{{\bf h}_5}(x^2y^3)^{{\bf h}_6}
(xy^2)^{{\bf h}_7}(xy^3)^{{\bf h}_8}y^{{\bf h}_9}
= 
y^{{\bf h}_9}(xy^3)^{{\bf h}_8}(xy^2)^{{\bf h}_7}
(x^2y^3)^{{\bf h}_6}(xy)^{{\bf h}_5}x^{{\bf h}_4}
\Fm^{-1}_{456789}
\end{align*}
and multiply it to (\ref{kanon1}) from the left.
The result reads
\begin{equation}\label{kanon15}
\begin{split}
&\Fm_{456789}^{-1}\bigl(x^{{\bf h}_4}L_{1_1 2_1 4} \cdots L_{1_n 2_n 4}\bigr)
\bigl((xy)^{{\bf h}_5}J_{1_1 3_1 2_1 5} \cdots J_{1_n 3_n 2_n 5}\bigr)
\bigl((x^2y^3)^{{\bf h}_6}L_{2_1 3_1 6} \cdots L_{2_n 3_n 6}\bigr)\\
&\quad\times 
\bigl((xy^2)^{{\bf h}_7}J_{2_1 1_1 3_1 7} \cdots J_{2_n 1_n 3_n 7}\bigr)
\bigl((xy^3)^{{\bf h}_8}L_{3_1 1_1 8} \cdots L_{3_n 1_n 8}\bigr)
\bigl(y^{{\bf h}_9}J_{3_1 2_1 1_1 9} \cdots J_{3_n 2_n 1_n 9}\bigr) 
\Fm_{456789}\\
&=
\bigl(y^{{\bf h}_9}J_{2_1 3_1 1_1 9} \cdots J_{2_n 3_n 1_n 9}\bigr)
\bigl((xy^3)^{{\bf h}_8}L_{1_1 3_1 8} \cdots L_{1_n 3_n 8}\bigr)
\bigl((xy^2)^{{\bf h}_7}J_{1_1 2_1 3_1 7} \cdots J_{1_n 2_n 3_n 7}\bigr)\\
&\quad\times
\bigl((x^2y^3)^{{\bf h}_6}L_{3_1 2_1 6} \cdots L_{3_n 2_n 6}\bigr)
\bigl((xy)^{{\bf h}_5}J_{3_1 1_1 2_1 5} \cdots J_{3_n 1_n 2_n 5}\bigr)
\bigl(x^{{\bf h}_4}L_{2_1 1_1 4} \cdots L_{2_n 1_n 4}\bigr).
\end{split}
\end{equation}

\subsection{Trace reduction}

Taking the trace of (\ref{kanon15}) over 
$\overset{4}{F}_{q^3}\otimes
\overset{5}{F}_{q} \otimes
\overset{6}{F}_{q^3} \otimes
\overset{7}{F}_{q}\otimes
\overset{8}{F}_{q^3} \otimes
\overset{9}{F}_{q}$,
we obtain
\begin{equation}\label{kanon2}
\begin{split}
&\mathrm{Tr}_4\bigl(x^{{\bf h}_4} 
L_{1_1 2_1 4} \cdots L_{1_n 2_n 4}\bigr)
\mathrm{Tr}_5\bigl((xy)^{{\bf h}_5}
J_{1_1 3_1 2_1 5} \cdots J_{1_n 3_n 2_n 5} \bigr) 
\mathrm{Tr}_6\bigl((x^2y^3)^{{\bf h}_6} 
L_{2_1 3_1 6} \cdots L_{2_n 3_n 6} \bigr)\\
&\times
\mathrm{Tr}_7\bigl((xy^2)^{{\bf h}_7}
J_{2_1 1_1 3_1 7} \cdots J_{2_n 1_n 3_n 7}\bigr)
\mathrm{Tr}_8\bigl((xy^3)^{{\bf h}_8}
L_{3_1 1_1 8} \cdots L_{3_n 1_n 8}\bigr)
\mathrm{Tr}_9\bigl(y^{{\bf h}_9}
J_{3_1 2_1 1_1 9} \cdots J_{3_n 2_n 1_n 9}\bigr)\\
&= \mathrm{Tr}_9\bigl(y^{{\bf h}_9}
J_{2_1 3_1 1_1 9} \cdots J_{2_n 3_n 1_n 9}\bigr)
\mathrm{Tr}_8\bigl((xy^3)^{{\bf h}_8}
L_{1_1 3_1 8} \cdots L_{1_n 3_n 8}\bigr)
\mathrm{Tr}_7\bigl((xy^2)^{{\bf h}_7}
J_{1_1 2_1 3_1 7} \cdots J_{1_n 2_n 3_n 7}\bigr)\\
& \times
\mathrm{Tr}_6\bigl((x^2y^3)^{{\bf h}_6}
L_{3_1 2_1 6} \cdots L_{3_n 2_n 6}\bigr)
\mathrm{Tr}_5\bigl((xy)^{{\bf h}_5}
J_{3_1 1_1 2_1 5} \cdots J_{3_n 1_n 2_n 5}\bigr)
\mathrm{Tr}_4\bigl(x^{{\bf h}_4}
L_{2_1 1_1 4} \cdots L_{2_n 1_n 4}\bigr).
\end{split}
\end{equation}
Here $\mathrm{Tr}_4(\cdots), \mathrm{Tr}_6(\cdots), \mathrm{Tr}_8(\cdots)$  
are identified with $R^{\mathrm{tr}}(z)$ in (\ref{obata1}). 
The other factors emerging from $J$ have the form
\begin{align}\label{sae}
G^{\mathrm{tr}}_{1 2 3}(z) 
= \kappa^{\mathrm{tr}}(z)
\mathrm{Tr}_a\bigl(z^{{\bf h}_a}
J_{1_1 2_1 3_1 a} \cdots J_{1_n 2_n 3_n a}\bigr)
\in \mathrm{End}(
\overset{1}{\bf V} \otimes \overset{2}{\bf V} 
\otimes \overset{3}{\bf V}),
\end{align}
where 
$\overset{k}{\bf V} = 
\overset{k_1}{V} \otimes \cdots \otimes \overset{k_n}{V}
\simeq (\C^2)^{\otimes n}$ as before.
The trace is taken over $\overset{a}{F}_q$ and evaluated 
by means of (\ref{ngh1}) and $(\ref{yuk1})|_{q\rightarrow q^{1/3}}$.
The scalar $\kappa^{\mathrm{tr}}(z)$ will be specified in (\ref{prpr}).
Now the relation (\ref{kanon2}) is rephrased as
\begin{equation}\label{misakiB}
\begin{split}
&R^{\mathrm{tr}}_{1 2}(x)
G^{\mathrm{tr}}_{1 3 2}(xy) 
R^{\mathrm{tr}}_{2 3}(x^2y^3)
G^{\mathrm{tr}}_{2 1 3}(xy^2)
R^{\mathrm{tr}}_{3 1}(xy^3)
G^{\mathrm{tr}}_{3 2 1}(y)\\
&=
G^{\mathrm{tr}}_{2 3 1}(y)
R^{\mathrm{tr}}_{1 3}(xy^3)
G^{\mathrm{tr}}_{1 2 3}(xy^2)
R^{\mathrm{tr}}_{3 2}(x^2y^3)
G^{\mathrm{tr}}_{3 1 2}(xy)
R^{\mathrm{tr}}_{2 1}(x).
\end{split}
\end{equation}
Thus the pair $(R^{\mathrm{tr}}(z), G^{\mathrm{tr}}(z))$ 
yields a solution to the $G_2$ reflection equation (\ref{reika}).

\subsection{Boundary vector reduction}

Set
\begin{align}
\langle \xi | = \sum_{m \ge 0} \frac{\langle m|}{(q)_m} \in 
F^\ast_{q},\qquad
|\xi\rangle = \sum_{m \ge 0} \frac{|m\rangle}{(q)_m}
\in F_{q},
\label{bve2}
\end{align}
which are formally the boundary vectors (\ref{bve}) with $q^3$ replaced by $q$.
Supported by computer experiments we conjecture\footnote{The two relations 
in (\ref{syki}) are actually equivalent due to the right property in (\ref{fpro}).} 
\begin{equation}\label{syki}
\begin{split}
(\langle \chi| \otimes 
\langle \xi | \otimes 
\langle \chi | \otimes
\langle \xi| \otimes 
\langle \chi | \otimes
\langle \xi| ) \Fm &= \langle \chi| \otimes 
\langle \xi | \otimes 
\langle \chi | \otimes
\langle \xi|\otimes 
\langle \chi | \otimes
\langle \xi| ,
\\
\Fm (|\chi\rangle \otimes
|\xi \rangle \otimes
|\chi \rangle \otimes
|\xi\rangle \otimes
|\chi \rangle \otimes
|\xi\rangle)
&= |\chi\rangle \otimes
|\xi \rangle \otimes
|\chi \rangle \otimes
|\xi\rangle \otimes
|\chi \rangle \otimes
|\xi\rangle,
\end{split}
\end{equation}
where $\langle \chi |$ and $|\chi\rangle$ are
defined in (\ref{bve}).
Sandwich the relation (\ref{kanon15}) 
between the bra vector
$\langle \overset{4}{\chi}| \otimes 
\langle \overset{5}{\xi}| \otimes 
\langle \overset{6}{\chi}| \otimes
\langle \overset{7}{\xi}|\otimes 
\langle \overset{8}{\chi}| \otimes
\langle \overset{9}{\xi}|$ 
and the ket vector 
$|\overset{4}{\chi}\rangle \otimes
|\overset{5}{\xi}\rangle \otimes
|\overset{6}{\chi}\rangle \otimes
|\overset{7}{\xi} \rangle\otimes
|\overset{8}{\chi}\rangle \otimes
|\overset{9}{\xi} \rangle$.
Thanks to (\ref{syki}) and the left relation in (\ref{fpro}), 
the result becomes
\begin{equation}\label{sizka0}
\begin{split}
&\langle \overset{4}{\chi}| 
x^{{\bf h}_4}L_{1_1 2_1 4} \cdots L_{1_n 2_n 4}|\overset{4}{\chi}\rangle 
\langle \overset{5}{\xi}| 
(xy)^{{\bf h}_5}J_{1_1 3_1 2_1 5} \cdots J_{1_n 3_n 2_n 5}|\overset{5}{\xi}\rangle
\langle \overset{6}{\chi}| 
(x^2y^3)^{{\bf h}_6}L_{2_1 3_1 6} \cdots L_{2_n 3_n 6}|\overset{6}{\chi}\rangle\\
&\quad\times 
\langle \overset{7}{\xi}| 
(xy^2)^{{\bf h}_7}J_{2_1 1_1 3_1 7} \cdots J_{2_n 1_n 3_n 7}|\overset{7}{\xi}\rangle
\langle \overset{8}{\chi}| 
(xy^3)^{{\bf h}_8}L_{3_1 1_1 8} \cdots L_{3_n 1_n 8}|\overset{8}{\chi}\rangle
\langle \overset{9}{\xi}| 
y^{{\bf h}_9}J_{3_1 2_1 1_1 9} \cdots J_{3_n 2_n 1_n 9}|\overset{9}{\xi}\rangle 
\\
&=
\langle \overset{9}{\xi}| 
y^{{\bf h}_9}J_{2_1 3_1 1_1 9} \cdots J_{2_n 3_n 1_n 9}|\overset{9}{\xi}\rangle
\langle \overset{8}{\chi}| 
(xy^3)^{{\bf h}_8}L_{1_1 3_1 8} \cdots L_{1_n 3_n 8}|\overset{8}{\chi}\rangle
\langle \overset{7}{\xi}| 
(xy^2)^{{\bf h}_7}J_{1_1 2_1 3_1 7} \cdots J_{1_n 2_n 3_n 7}|\overset{7}{\xi}\rangle\\
&\quad\times
\langle \overset{6}{\chi}| 
(x^2y^3)^{{\bf h}_6}L_{3_1 2_1 6} \cdots L_{3_n 2_n 6}|\overset{6}{\chi}\rangle
\langle \overset{5}{\xi}| 
(xy)^{{\bf h}_5}J_{3_1 1_1 2_1 5} \cdots J_{3_n 1_n 2_n 5}|\overset{5}{\xi}\rangle
\langle \overset{4}{\chi}| 
x^{{\bf h}_4}L_{2_1 1_1 4} \cdots L_{2_n 1_n 4}|\overset{4}{\chi}\rangle.
\end{split}
\end{equation}
The factors 
$\langle \chi|(\cdots)| \chi\rangle$ involving $L$ 
are identified with $R^{\mathrm{bv}}(z)$ in (\ref{obata2}).
The other factors emerging from $J$ have the form
\begin{align}\label{sizka}
G^{\mathrm{bv}}_{123}(z) = \kappa^{\mathrm{bv}}(z)
\langle \overset{a}{\xi}|
z^{{\bf h}_a}
J_{1_1 2_1 3_1 a}\cdots J_{1_n 2_n 3_n a}
|\overset{a}{\xi}\rangle \in 
\mathrm{End}(\overset{1}{\bf V}\otimes 
\overset{2}{\bf V}\otimes \overset{3}{\bf V}),
\end{align}
where the scalar $\kappa^{\mathrm{bv}}(z)$ will be specified in (\ref{prpr}).
In terms of (\ref{sizka}) and (\ref{s11}), 
the relation (\ref{sizka0}) is stated as 
\begin{equation}\label{aoi}
\begin{split}
&R^{\mathrm{bv}}_{1 2}(x)
G^{\mathrm{bv}}_{1 3 2}(xy) 
R^{\mathrm{bv}}_{2 3}(x^2y^3)
G^{\mathrm{bv}}_{2 1 3}(xy^2)
R^{\mathrm{bv}}_{3 1}(xy^3)
G^{\mathrm{bv}}_{3 2 1}(y)\\
&=
G^{\mathrm{bv}}_{2 3 1}(y)
R^{\mathrm{bv}}_{1 3}(xy^3)
G^{\mathrm{bv}}_{1 2 3}(xy^2)
R^{\mathrm{bv}}_{3 2}(x^2y^3)
G^{\mathrm{bv}}_{3 1 2}(xy)
R^{\mathrm{bv}}_{2 1}(x).
\end{split}
\end{equation}
Thus the pair $(R^{\mathrm{bv}}(z), G^{\mathrm{bv}}(z))$ 
provides another solution to the $G_2$ reflection equation (\ref{reika}).
 
\subsection{\mathversion{bold}Basic properties of 
$G^{\mathrm {tr}}(z)$ and $G^{\mathrm {bv}}(z)$}\label{ss:bpg}

The construction (\ref{sae}) and (\ref{sizka}) 
imply the matrix product formula for each element as
\begin{align}
G(z) (|\alb\rangle \otimes |\beb \rangle \otimes  |\gab\rangle)
&= \sum_{\lab, \mub, \nub \in \{0,1\}^n}
G(z)^{\lab, \mub, \nub}_{\alb, \beb, \gab}\,
|\lab\rangle \otimes  |\mub \rangle  \otimes |\nub\rangle
\qquad(G = G^{\mathrm{tr}}, G^{\mathrm{bv}}),
\label{misato}\\
G^{\mathrm{tr}}(z)^{\lab, \mub, \nub}_{\alb, \beb, \gab}
& = \kappa^{\mathrm{tr}}(z)
\mathrm{Tr}\bigl(z^{{\bf h}}
J^{\lambda_1, \mu_1,\nu_1}_{\alpha_1, \beta_1, \gamma_1}
\cdots 
J^{\lambda_n, \mu_n,\nu_n}_{\alpha_n, \beta_n, \gamma_n}
\bigr),
\label{ana1}\\
G^{\mathrm{bv}}(z)^{\lab, \mub, \nub}_{\alb, \beb, \gab}
& = \kappa^{\mathrm{bv}}(z)
\langle \xi | z^{{\bf h}}
J^{\lambda_1, \mu_1,\nu_1}_{\alpha_1, \beta_1, \gamma_1}
\cdots 
J^{\lambda_n, \mu_n,\nu_n}_{\alpha_n, \beta_n, \gamma_n}
|\xi\rangle
\label{ana2}
\end{align}
in terms of $J^{\lambda, \mu,\nu}_{\alpha, \beta, \gamma}$ 
specified in (\ref{Jdef})--(\ref{spd}).
From (\ref{yumi1}) and (\ref{yumi2}) one can show
\begin{align}\label{misato2}
G^{\mathrm{tr}}(z)^{\lab, \mub, \nub}_{\alb, \beb, \gab} = 0
\quad\text{unless}\;\;
\alb+\beb = \lab + \mub \in \Z^n\;\;\text{and}\;\;
n+|\beb|-|\gab| = |\mub|+|\nub|
\end{align}
or equivalently the direct sum decomposition:
\begin{equation}\label{misato3}
\begin{split}
G^{\mathrm{tr}}(z) = \bigoplus_{l,m,k} 
G^{\mathrm{tr}}(z)_{l,m,k},\qquad
G^{\mathrm{tr}}(z)_{l,m,k}: 
\Vb_l\otimes \Vb_m \otimes \Vb_k 
\rightarrow \bigoplus_{k'}\Vb_{l+k+k'-n}\otimes 
\Vb_{m-k-k'+n} \otimes \Vb_{k'},
\end{split}
\end{equation}
where the sums extend over $l,m,k,k' \in [0,n]$
such that the indices $l+k+k'-n$ and 
$m-k-k'+n$ also belong to $[0,n]$.

The trace (\ref{ana1}) is evaluated by means of (\ref{yuk1}) with 
$q^3$ replaced by $q$.
The quantity $\langle \xi|(\cdots)|\xi\rangle$  in (\ref{ana2}) 
is calculated from (\ref{ngh1}) and 
\begin{equation}\label{lin}
\langle \xi|z^{\bf h} (\apm)^j \ok^m|\xi\rangle 
= q^{\frac{m}{2}}(-q;q)_j
\frac{(-q^{j+m+1}z;q)_\infty}{(q^mz;q)_\infty}\times
\begin{cases}
z^j, \\
q^{mj},
\end{cases}
\end{equation}
which is easily derived by only using the elementary identity
$\sum_{j\ge 0}\frac{(w;q)_j}{(q;q)_j}z^j 
= \frac{(w z;q)_\infty}{(z;q)_\infty}$.
We choose the normalization factors in (\ref{ana1}) and (\ref{ana2}) as
\begin{align}\label{prpr}
\kappa^{\mathrm{tr}}(z)=1,\qquad
\kappa^{\mathrm{bv}}(z)=
\frac{(z;q)_\infty}{(-qz;q)_\infty}.
\end{align}
Then all the matrix elements (\ref{ana1}) and (\ref{ana2}) become 
rational in $z$ and $q^{\hf}$.
For instance we have
\begin{align}
G^{\mathrm{tr}}(z)
(|{\bf e}_{[1,l]}\rangle
\otimes |{\bf e}_{[1,m]}\rangle \otimes |{\bf 0}\rangle)
&= \frac{(q^{\hf})^{m-l+n}}{1-zq^{m-l+n}}
|{\bf e}_{[1,l]}\rangle
\otimes |{\bf e}_{[1,m]}\rangle \otimes |{\bf 1}\rangle + \cdots\quad (l\le m),
\\
G^{\mathrm{tr}}(z)
(|{\bf e}_{[1,l]}\rangle
\otimes |{\bf e}_{[1,m]}\rangle \otimes |{\bf 1}\rangle)
&= \frac{(-q^{\hf})^{l-m+n}}{1-zq^{l-m+n}}
|{\bf e}_{[1,l]}\rangle
\otimes |{\bf e}_{[1,m]}\rangle \otimes |{\bf 0}\rangle + \cdots\quad (l\ge m),
\\
G^{\mathrm{bv}}(z)
(|{\bf e}_{[1,l]}\rangle
\otimes |{\bf e}_{[1,m]}\rangle \otimes |{\bf 0}\rangle)
&= q^{\frac{m-l+n}{2}}\frac{(z;q)_{m-l+n}}{(-qz;q)_{m-l+n}}
|{\bf e}_{[1,l]}\rangle
\otimes |{\bf e}_{[1,m]}\rangle \otimes |{\bf 1}\rangle + \cdots\quad (l\le m),
\\
G^{\mathrm{bv}}(z)
(|{\bf e}_{[1,l]}\rangle
\otimes |{\bf e}_{[1,m]}\rangle \otimes |{\bf 1}\rangle)
&= (-q^\hf)^{l-m+n}
\frac{(z;q)_{l-m+n}}{(-qz;q)_{l-m+n}}
|{\bf e}_{[1,l]}\rangle
\otimes |{\bf e}_{[1,m]}\rangle \otimes |{\bf 0}\rangle + \cdots\quad (l\ge m),
\end{align}
where the symbol ${\bf e}_{[1,k]}$ was defined after (\ref{askS1}) and 
$|{\bf 1}\rangle = |{\bf e}_{[1,n]}\rangle = |1,1,\ldots, 1\rangle$.

\begin{remark} 
Using (\ref{nami}) and (\ref{misato3}) it is easy to see that 
the both sides of (\ref{misakiB}) applied to
$\Vb_s\otimes \Vb_t \otimes \Vb_u$ 
generate the space
$\bigoplus_{t'}
\Vb_{t-s+t'} \otimes \Vb_{t'} \otimes \Vb_{2n-u-t-t'}$.
There are three $R^{\mathrm{tr}}(z)$'s on each side of (\ref{misakiB}).
One can check that changing their normalization as 
$R^{\mathrm{tr}}_{l,m}(z) \rightarrow \phi_{l,m}(z)R^{\mathrm{tr}}_{l,m}(z)$
depending on $l,m$ in (\ref{nami}) keeps  (\ref{misakiB}) valid  
for any function  
$\phi_{l,m}(z)$ of the form 
$\tilde{\phi}_{l-m}(z)$.
In fact the both sides acquire the common overall factor
${\tilde \phi}_{s+u-n}(xy^3)
{\tilde \phi}_{t+u-n}(x^2y^3)
{\tilde \phi}_{t-s}(x)$ under the change.
Our $\varrho^{\mathrm{tr}}_{l,m}(z)$ (\ref{askS}) is of this form,
hence (\ref{misakiB}) remains valid 
despite the ``mixing"  of weights (or indices) in (\ref{misato3}).
\end{remark}

\section{Summary}\label{sec:end}

We have studied the $G_2$ reflection equation (\ref{reika})
which is a natural $G_2$ analogue of the 
Yang-Baxter and the reflection equations corresponding to the
$A_2$ and $B_2/C_2$ Coxeter relations, respectively.
It describes the three particle scattering/reflections
whose world-lines form a Pappus configuration.
We introduced the quantized $G_2$ reflection equation (\ref{qre}).
It is a $q$-boson valued $G_2$ reflection equation that holds up to conjugation.
We gave a solution to it in Theorem \ref{th:sol}
by exploiting a connection to the quantized coordinate ring 
$A_q(G_2)$ \cite{KOY}.
From the concatenation of the solution we have constructed 
matrix product solutions to the original $G_2$ reflection equation
$(R^{\mathrm{tr}}(z), G^{\mathrm{tr}}(z))$ in (\ref{misakiB}) and 
$(R^{\mathrm{bv}}(z), G^{\mathrm{bv}}(z))$ in (\ref{aoi}), where 
the latter assumes the conjecture (\ref{syki}).
The $R$ and $G$ matrices are linear operators on $\Vb \otimes \Vb$
and $\Vb \otimes \Vb \otimes \Vb$ with $\Vb \simeq (\C^2)^{\otimes n}$
and trigonometric in the spectral parameter.
The special three particle event characteristic to the $G_2$ theory 
is encoded in $G^{\mathrm{tr}}(z), G^{\mathrm{bv}}(z)$ 
whereas the companion $R$ matrices 
$R^{\mathrm{tr}}(z), R^{\mathrm{bv}}(z)$  for the two particle scattering 
are the known ones 
for the antisymmetric tensor representations of $U_p(A^{(1)}_{n-1})$
and the spin representations of $U_p(D^{(2)}_{n+1})$ with $p^2=-q^{-3}$.

\appendix 

\section{Correspondence between (\ref{minami}) and (\ref{kkna})}
\label{app:ind}
We indicate the equivalence of (\ref{minami}) with  $(a,b,c,i,j,k) \in \{0,1\}^6$ 
and (\ref{kkna}) with $(i,j) \rightarrow (i',j')\in \{1,\ldots, 7\}^2$ as $(abcijk; i'j')$.
In this notation the example treated in (\ref{eqd16}) and 
(\ref{eq16}) is $(111100; 16)$.
There is no need to take linear combinations etc of the equations and 
the equivalence literally means the same equation up to an overall constant.
\begin{align*}
&\quad({0 0 0 0 0 0}; {7 7} ), \quad ({0 0 0 0 0 1}; {7 4} ),
\quad ({0 0 0 0 1 0}; {7 5} ), \quad ({0 0 0 0 1 1}; {7 2} ), 
\quad ({0 0 0 1 0 0}; {7 6} ), \quad ({0 0 0 1 0 1}; {7 3} ),
\\ & \quad ({0 0 0 1 1  0}; {7 4} ), \quad ({0 0 0 1 1 1}; {7 1} ), 
\quad ({0 0 1 0 0 0}; {4 7} ), \quad ({0 0 1 0 0 1}; {4 4} ), 
\quad ({0 0 1 0 1 0}; {4 5} ), \quad ({0 0 1 0 1 1}; {4 2} ),
 \\ &\quad ({0 0 1 1 0  0}; {4 6} ), \quad ({0 0 1 1 0 1}; {4 3} ), 
\quad ({0 0 1 1 1 0}; {4 4} ), \quad ({0 0 1 1 1 1}; {4 1} ), 
\quad ({0 1 0 0 0 0}; {5 7} ), \quad ({0 1 0 0 0 1}; {5 4} ), 
\\ &\quad ({0 1 0 0 1 0}; {5 5} ), \quad ({0 1 0 0 1 1}; {5 2} ), 
\quad ({0 1 0 1 0 0}; {5 6} ), \quad ({0 1 0 1 0 1}; {5 3} ), 
\quad ({0 1 0 1 1 0}; {5 4} ), \quad ({0 1 0 1 1 1}; {5 1} ), 
\\ &\quad ({0 1 1 0 0 0}; {2 7} ), \quad ({0 1 1 0 0 1}; {2 4} ), 
\quad ({0 1 1 0 1 0}; {2 5} ), \quad ({0 1 1 0 1 1}; {2 2} ), 
\quad ({0 1 1 1 0 0}; {2 6} ), \quad ({0 1 1 1 0 1}; {2 3} ), 
\\ &\quad ({0 1 1 1 1 0}; {2 4} ), \quad ({0 1 1 1 1 1}; {2 1} ), 
\quad ({1 0 0 0 0 0}; {6 7} ), \quad ({1 0 0 0 0 1}; {6 4} ), 
\quad ({1 0 0 0 1 0}; {6 5} ), \quad ({1 0 0 0 1 1}; {6 2} ), 
\\ &\quad ({1 0 0 1 0 0}; {6 6} ), \quad ({1 0 0 1 0 1}; {6 3} ), 
\quad ({1 0 0 1 1 0}; {6 4} ), \quad ({1 0 0 1 1 1}; {6 1} ), 
\quad ({1 0 1 0 0 0}; {3 7} ), \quad ({1 0 1 0 0 1}; {3 4} ), 
\\ &\quad ({1 0 1 0 1 0}; {3 5} ), \quad ({1 0 1 0 1 1}; {3 2} ), 
\quad ({1 0 1 1 0 0}; {3 6} ), \quad ({1 0 1 1 0 1}; {3 3} ), 
\quad ({1 0 1 1 1 0}; {3 4} ), \quad ({1 0 1 1 1 1}; {3 1} ), 
\\ &\quad ({1 1 0 0 0 0}; {4 7} ), \quad ({1 1 0 0 0 1}; {4 4} ), 
\quad ({1 1 0 0 1 0}; {4 5} ), \quad ({1 1 0 0 1 1}; {4 2} ), 
\quad ({1 1 0 1 0 0}; {4 6} ), \quad ({1 1 0 1 0 1}; {4 3} ), 
\\ &\quad ({1 1 0 1 1 0}; {4 4} ), \quad ({1 1 0 1 1 1}; {4 1} ), 
\quad ({1 1 1 0 0 0}; {1 7} ), \quad ({1 1 1 0 0 1}; {1 4} ), 
\quad ({1 1 1 0 1 0}; {1 5} ), \quad ({1 1 1 0 1 1}; {1 2} ), 
\\ &\quad ({1 1 1 1 0 0}; {1 6} ), \quad ({1 1 1 1 0 1}; {1 3} ), 
\quad ({1 1 1 1 1 0}; {1 4} ), \quad ({1 1 1 1 1 1}; {1 1}).
\end{align*}
Note for instance that $({0 0 0 0 0 1}; {7 4} )$
and $({0 0 0 1 1  0}; {7 4} )$ imply that not all of the 
quantized reflection equations (\ref{minami}) are independent.

\section{Examples}\label{app:ex}

Recall the notation $|\alb \rangle \in 
\Vb=V^{\otimes n} \simeq (\C^2)^{\otimes n}$ declared in the beginning of 
Section \ref{ss:bss}.
We write $|\alb \rangle \otimes |\beb\rangle \in \Vb \otimes \Vb$
and 
$|\alb \rangle \otimes |\beb\rangle \otimes |\gab\rangle 
\in \Vb \otimes \Vb \otimes \Vb$
with $\alpha=(\alpha_1,\ldots, \alpha_n) \in \{0,1\}^n$ etc as
$|\alpha_1\ldots \alpha_n, \beta_1 \ldots \beta_n\rangle$
and 
$|\alpha_1\ldots \alpha_n, \beta_1 \ldots \beta_n, \gamma_1\ldots \gamma_n\rangle$,
respectively.

\subsection{\mathversion{bold}$R^{\mathrm{tr}}(z)$ 
and $G^{\mathrm{tr}}(z)$ with $n=1$}

$R^{\mathrm{tr}}(z)$ is given by 
$|i,j\rangle \mapsto (-1)^{i(1-j)}|i,j\rangle\, (i,j=0,1)$, which  
is almost the identity.
On the other hand 
$G^{\mathrm{tr}}(z)$ is nontrivial even for $n=1$.
Its action on $\Vb^{\otimes 3}= V^{\otimes 3}$ is given by 
\begin{align*}
|0,0,0\rangle &\mapsto \frac{q^\hf |0,0,1\rangle}{1-q z},
\quad
|0,0,1\rangle \mapsto -\frac{q^\hf |0,0,0\rangle}{1-q z},
\quad
|0,1,0\rangle \mapsto \frac{q |0,1,1\rangle}{1-q^2 z},\\
|0,1,1\rangle &\mapsto 
-\frac{u_1u_3(q^2-z)|0,1,0\rangle}{r(1-z)(1-q^2z)}
-\frac{u_3u_4(q^2-z)|1,0,1\rangle}{r(1-z)(1-q^2z)},\\
|1,0,0\rangle &\mapsto 
-\frac{u_1u_2(q^2-z)|0,1,0\rangle}{r(1-z)(1-q^2z)}
-\frac{u_2u_4(q^2-z)|1,0,1\rangle}{r(1-z)(1-q^2z)},\\
|1,0,1\rangle &\mapsto \frac{q |1,0,0\rangle}{1-q^2 z},
\quad
|1,1,0\rangle \mapsto \frac{q^\hf |1,1,1\rangle}{1-q z},
\quad
|1,1,1\rangle \mapsto -\frac{q^\hf |1,1,0\rangle}{1-q z},
\end{align*}
where $r, u_1, u_2, u_3, u_4$ are to obey (\ref{rdef}).
The two kinds of the denominators $1-qz$ and $1-q^2z$ 
originate in $J^{0,0,1}_{0,0,0}=\ok$
and $J^{0,1,1}_{0,1,0}=\ok^2$.

\subsection{\mathversion{bold}$R^{\mathrm{tr}}(z)$ 
and $G^{\mathrm{tr}}(z)$ with $n=2$}

$R^{\mathrm{tr}}_{l,m}(z)\,(0 \le l, m \le 2)$ is the identity except 
$R^{\mathrm{tr}}_{1,0}(z)=-\mathrm{id}$,
$R^{\mathrm{tr}}_{2,1}(z)=-\mathrm{id}$ and 
$R^{\mathrm{tr}}_{1,1}(z)$.
The last one $R^{\mathrm{tr}}_{1,1}(z)$ is given by 
\begin{align*}
|ij,ij\rangle &\mapsto |ij,ij\rangle\quad(i=1-j=0,1),\\
|01,10\rangle &\mapsto 
-\frac{q^3(1-z)|01,10\rangle}{1-q^6z}+
\frac{(1-q^6)z|10,01\rangle}{1-q^6z},\\
|10,01\rangle &\mapsto 
\frac{(1-q^6)|01,10\rangle}{1-q^6z}
-\frac{q^3(1-z)|10,01\rangle}{1-q^6z},
\end{align*}
which is a six-vertex. 
As for $G^{\mathrm{tr}}(z)$, it is too lengthy to present all the data.
So we give just a few examples.
\begin{align*}
|00,00,00\rangle &\mapsto 
\frac{q|00,00,11\rangle}{1-q^2z},\quad
|00,00,01\rangle \mapsto
\frac{(1-q^2)z|00,00,01\rangle}{(1-z)(1-q^2z)}
-\frac{q|00,00,10\rangle}{1-q^2z},\\
|00,10,11\rangle &\mapsto
\frac{q^\thf u_1u_3(q-z)|00,10,00\rangle}{r(1-qz)(1-q^3z)}
-\frac{q^\hf (1-q^2)u_3z|10,00,01\rangle}{(1-qz)(1-q^3z)}
+\frac{q^\thf u_3u_4(q-z)|10,00,10\rangle}{r(1-qz)(1-q^3z)},\\ 
|10,01,01\rangle &\mapsto
\frac{u_1^2 u_2 u_3 (q^4 + z - 2 q^2 z - 2 q^4 z + q^6 z + 
     q^2 z^2) |{{00}, {1 1}, {0 0}}\rangle)}{r^2 (1 - z) (1 -
      q^2 z) (1 - q^4 z)} 
\\&\quad
+ \frac{
 u_1 u_2 u_3 u_4 (q^4 + z - 2 q^2 z - 2 q^4 z + q^6 z + 
    q^2 z^2) |{{0 1}, {1 0}, {0 1}}\rangle}{r^2 (1 - z) (1 - 
    q^2 z) (1 - q^4 z)} 
\\&\quad
- \frac{q(1-q^2) u_2 u_3 |{{0 1}, {1 0}, {1 0}}\rangle}{(1 - q^2 z) (1 - 
    q^4 z)} 
- \frac{q(1-q^2) u_2 u_3 z |{{1 0}, {0 1}, {0 1}}\rangle}{(1 - q^2 z)(1 - 
    q^4 z)} 
\\&\quad
+ \frac{ u_1 u_2 u_3 u_4 (q^4 + z - 2 q^2 z - 2 q^4 z + q^6 z + 
    q^2 z^2) |{{1 0}, {0 1}, {1 0}}\rangle}{r^2 (1 - z) (1 - 
    q^2 z) (1 - q^4 z)}
\\&\quad 
+ \frac{ u_2 u_3 u_4^2 (q^4 + z - 2 q^2 z - 2 q^4 z + q^6 z + 
    q^2 z^2) |{{1 1}, {0 0}, {1 1}}\rangle}{r^2 (1 - z) (1 - 
    q^2 z) (1 - q^4 z)}.
\end{align*}

\subsection{\mathversion{bold}$R^{\mathrm{bv}}(z)$ 
and $G^{\mathrm{bv}}(z)$ with $n=1$}

$R^{\mathrm{bv}}(z)$ reduces to another six-vertex model:
\begin{align*}
|i,i\rangle &\mapsto |i,i\rangle\quad (i=0,1),\\
|0,1\rangle &\mapsto \frac{q^\thf(1-z)|0,1\rangle}{1+q^3z}
+\frac{(1+q^3)z|1,0\rangle}{1+q^3z},\qquad
|1,0\rangle \mapsto \frac{(1+q^3)|0,1\rangle}{1+q^3z}
-\frac{q^\thf(1-z)|1,0\rangle}{1+q^3z}.
\end{align*}
$G^{\mathrm{bv}}(z)$ is given by
\begin{align*}
|0,0,0\rangle &\mapsto
\frac{(1+q)z|0,0,0\rangle}{1+qz}+
\frac{q^\hf(1-z)|0,0,1\rangle}{1+qz},
\quad
|0,0,1\rangle \mapsto 
-\frac{q^\hf(1-z)|0,0,0\rangle}{1+qz}+
\frac{(1+q)|0,0,1\rangle}{1+qz},\\
|0,1,1\rangle & \mapsto
\frac{q^\thf(1+q)u_1(1-z)z|0,1,0\rangle}{(1+qz)(1+q^2z)}
+\frac{q(1-z)(1-qz)|0,1,1\rangle}{(1+qz)(1+q^2z)}\\
&+\frac{(1+q)(1+q^2)z^2|1,0,0\rangle}{(1+qz)(1+q^2z)}
+\frac{q^\thf(1+q)u_4(1-z)z|1,0,1\rangle}{(1+qz)(1+q^2z)},\\
|0,1,1\rangle &\mapsto
\frac{u_3(-q^2 + z + 2 q z + 2 q^2 z + q^3 z - q z^2)
(u_1|0,1,0\rangle+u_4|1,0,1\rangle)}{r(1+qz)(1+q^2z)}\\
&+
\frac{q^\hf(1+q)u_3(1-z)(|0,1,1\rangle-z|1,0,0\rangle)}{(1+qz)(1+q^2z)},\\
|1,0,0\rangle &\mapsto
\frac{ u_2 (-q^2 + z + 2 q z + 2 q^2 z + q^3 z - 
   q z^2)(u_1 |0, 1, 0\rangle+u_4|1,0,1\rangle)}{r(1 + q z) (1 + q^2 z)}\\
&+ \frac{q^\hf (1+q)u_2(1-z)(|0,1,1\rangle-z |1,0,0\rangle)}{(1+qz)(1+q^2z)},\\
|1,0,1\rangle &\mapsto
-\frac{q^\thf(1+q)u_1(1-z)|0,1,0\rangle}{(1+qz)(1+q^2z)}
+\frac{(1+q)(1+q^2)|0,1,1\rangle}{(1+qz)(1+q^2z)}\\
&+
\frac{q(1-z)(1-qz)|1,0,0\rangle}{(1+qz)(1+q^2z)}-
\frac{q^\thf(1+q)u_4(1-z)|1,0,1\rangle}{(1+qz)(1+q^2z)},\\
|1,1,0\rangle &\mapsto
\frac{(1+q)z|1,1,0\rangle}{1+qz}+
\frac{q^\hf(1-z)|1,1,1\rangle}{1+qz},\quad
|1,1,1\rangle \mapsto 
-\frac{q^\hf(1-z)|1,1,0\rangle}{1+qz}+\frac{(1+q)|1,1,1\rangle}{1+qz}.
\end{align*}

\section*{Acknowledgments}
The author thanks Ivan Cherednik,  
Yasushi Komori, Masato Okado and  Yasuhiko Yamada for communications 
and comments.
This work is supported by 
Grants-in-Aid for Scientific Research 
No.~18H01141 from JSPS.

\end{document}